\documentclass[a4paper,11pt]{article}
\usepackage{jcappub}
\usepackage[T1]{fontenc}
\usepackage[caption=false]{subfig}
\usepackage{hyperref}
\usepackage{amsmath}
\usepackage{mathtools}
\usepackage{tablefootnote}
\usepackage{soul}
\usepackage{ulem}
\usepackage{color}
\usepackage{float}
\usepackage{multirow}
\usepackage{comment}
\usepackage{makecell}
\usepackage[dvipsnames,svgnames,table]{xcolor}
\usepackage{amssymb}
\usepackage[makeroom]{cancel}
\usepackage{graphicx}
\usepackage{subcaption}
\usepackage{booktabs}
\usepackage{xspace}

\newcommand{\hgl}{\texttt{HGL}}
\newcommand{\cosmo}{\theta_{\mathcal{C}}}
\newcommand{\astro}{\theta_{\mathcal{A}}}
\newcommand{\data}{\mathbf{d}}
\newcommand{\params}{\boldsymbol{\theta}}

\title{Field-level multi-tracers simulation-based inference of cosmological parameters from 3D maps}

\author[a,b]{Giulio Scelfo} 
\author[a]{Satvik Mishra}
\author[a]{Mauro Rigo}
\author[a,d,e]{Roberto Trotta} 
\author[a,b,c,d]{Matteo Viel}

\emailAdd{gscelfo@sissa.it}

\affiliation[a]{SISSA, Via Bonomea 265, 34136 Trieste, Italy \& INFN Sezione di Trieste}
\affiliation[b]{INFN - National Institute for Nuclear Physics, Via Valerio 2, I-34127 Trieste, Italy}
\affiliation[c]{IFPU - Institute for Fundamental Physics of the Universe, Via Beirut 2, 34014 Trieste, Italy}
\affiliation[d]{ICSC -- Centro Nazionale di Ricerca in High Performance Computing, Big Data e Quantum Computing, Via Magnanelli 2, Bologna, Italy}
\affiliation[e]{Physics Department, Blackett Lab, Imperial College London, Prince Consort Road, London SW7 2AZ, UK}

\abstract{Extracting maximum cosmological information from current and upcoming large-scale structure data requires going beyond summary statistics as currently used in likelihood-based inference. Simulation-Based Inference (SBI) promises to enable the exploitation of field-level information and the rich physics of modern hydrodynamical simulations. We develop a proof-of-concept SBI pipeline to explore its potential to constrain the cosmological parameters $\{\Omega_{\rm m}, \sigma_8\}$ from galaxy number counts, neutral hydrogen (HI) intensity mapping and their combination. We use neural emulators trained on full hydrodynamical simulations to generate galaxy and HI maps from fast, approximate dark matter simulations. Combined with neural posterior estimation, this enables the estimation of cosmological parameters while marginalizing over astrophysical effects. We perform inference both on the power spectrum and on representations derived from field-level 2D or 3D maps, comparing results from each probe and the combination of both tracers, and assessing the impact of data compression and multi-tracers information on cosmological constraints. 

Combining galaxy and HI fields improves constraints with respect to single-tracer cases by a factor 2 to 7 in terms of a Figure of Merit describing the joint precision on cosmological parameters, depending on the tracer/configuration. Moving from summary statistics to field-level inference leads to a consistent gain in constraining power of about a factor 3, with 3D maps providing the most precise and well-calibrated posteriors. This gain in precision is robust even when astrophysical parameters are marginalized over. Further developments (including realistic survey effects and improvements in emulators' faithfulness) will enable the application of this analysis pipeline to upcoming surveys. 
}

\begin{document}
\maketitle

\section{Introduction}
The current and upcoming generation of large-scale structure (LSS) surveys, such as Euclid~\cite{mellier2025euclid}, the Vera Rubin Observatory~\cite{LSST}, SKAO~\cite{Braun2015:ska} and DESI~\cite{aghamousa2016desi}, will deliver data that encode valuable cosmological information in the full non-Gaussian structure of observable fields, beyond what can be extracted from compressed, hand-crafted summaries like two-point statistics. Due to the volume of such data sets, statistical uncertainties will become increasingly subdominant, while systematic effects and modeling accuracy will become the dominant limiting factors. In this context, extracting cosmological information from observables such as galaxy clustering and neutral hydrogen (HI) intensity mapping requires an accurate forward modeling of baryonic observables in highly non-linear structures across a wide range of scales. This is challenging: the mapping between cosmological and astrophysical parameters and the resulting observables is governed by complex, non-linear gravitational dynamics and baryonic processes, which can only be fully captured through numerical simulations. The development of open cosmological simulations suites such as, e.g., \texttt{CAMELS}~\cite{camels}, Quijote~\cite{Quijote_sims} and DREAMS~\cite{Dreams_project} aims at building a solid foundation towards this goal.

High-fidelity simulations, particularly full hydrodynamical ones, are computationally expensive, often requiring millions of CPU hours to run, and therefore it is unfeasible to run them in sufficient numbers across the high-dimensional parameter space relevant for modern cosmology. This poses a fundamental obstacle for traditional inference pipelines, in which a Markov Chain Monte Carlo approach~\cite{brooks:mcmc,Trotta:2008qt} requires the evaluation of a likelihood function at every sampled point in parameter space. Given the impossibility of obtaining full hydrodynamical simulations covering the whole parameter space, the standard approach is to adopt approximate likelihoods, typically assumed to be Gaussian, defined in terms of a summary statistics, which are often limited to the power spectrum. While computationally tractable~\cite{Halofit}, these approaches are known to discard a significant fraction of the cosmological information encoded in the non-Gaussian features of the matter and tracer fields, especially on small and intermediate scales. For example,~\cite{Leclercq21} shows that, on a simplified analytic model based on lognormal fields, often used as an approximation to the late-time cosmological density field, standard approximations adopted to model a likelihood for correlation functions can cause significant biases, whereas field-level inference gives more accurate and precise posteriors, with gains increasing with non-Gaussianity. Similar conclusions have been reached by~\cite{Porqueres22} in regards of cosmic shear data. However, the exact magnitude of the gain of full-field analyses with respect to Gaussian summary-statistics likelihoods is context-dependent: crucial factors include the degree of non-Gaussianity and the scale considered~\cite{Hahn19, Leclercq21}, the characteristics of the instrument adopted and whether the summary statistic is actually close to sufficient (see e.g.,~\cite{Upham21, Brieden_2021}).

In the last few years, Simulation-Based Inference (SBI) has emerged as a powerful framework that circumvents the need for an explicit likelihood function (see e.g.~\cite{Cranmer20:SBI,lueckmann21:SBI} for reviews). Instead of specifying a likelihood, SBI relies on forward simulations and machine learning methods to learn the posterior distribution, the likelihood function or the likelihood-to-evidence ratio of parameters $\theta$ given the data $x$. This makes SBI particularly well-suited for cosmology, where numerical simulations are already central to modeling the non-linear regime of structure formation~\cite{angulo_large-scale_2022}, but where analytical likelihoods are often intractable. Early applications to weak-lensing maps showed that likelihood-free inference can combine forward modeling, realistic survey masks and noise, and neural compression of summary statistics~\cite{jeffrey_likelihood-free_2020}. 

SBI has already been demonstrated to be a powerful approach to inference in a range of cosmological and astrophysical applications. For example,~\cite{karchev2023simsimssimulationbasedsupernovaia} used neural classification for Bayesian model comparison in Type Ia supernova analyses, demonstrating that SBI can marginalize over thousands of latent variables. In cosmological field reconstruction,~\cite{list2023bayesian} introduced an SBI-based Bayesian framework applicable to generic non-differentiable forward simulators, enabling sampling from the posterior of the underlying field. In 21-cm cosmology,~\cite{Saxena_2024} applied truncated marginal neural ratio estimation to mock observations, showing that SBI can effectively handle complex foreground, beam, and noise effects while targeting marginal posteriors efficiently. In studies of galactic binaries with gravitational wave data,~\cite{Srinivasan:2025etu} demonstrated that SBI can recover population-level parameters bypassing individual sources' fits. In exoplanetary atmospheric retrieval,~\cite{Lueber_2025} obtained orders-of-magnitude efficiency increase with respect to nested sampling methods, enabling large-scale model comparison studies. More broadly, SBI is flexible enough to operate both on summary statistics, such as the traditional power spectrum and bispectrum (see e.g.~\cite{sinigaglia2026simulation}) or more advanced ones (see e.g.~\cite{gorbatchev2026}), and on full field-level data using realistic forward models (see e.g.~\cite{lemos2023simbig, tucci24}). Complementarily, recent field-level SBI methods have also targeted the reconstruction and fast posterior sampling of cosmological initial conditions \citep{savchenko2024mean, savchenko2025fast}.

Despite its appeal, SBI requires a large number of simulations to in order to train the inference model. Recent weak-lensing studies have benchmarked full-field approaches, emphasizing both the potential optimality of field-level inference and the large simulation budgets required to achieve sufficient generalization performance~\cite{Zeghal_2025}. This issue has also been studied in~\cite{bairagi2025simulationsneedsimulationbasedinference}, which showed that the number of simulations needed for neural information extraction can be substantially larger than in currently available simulation suites. To address this bottleneck, emulators have become an essential component of modern inference pipelines. Emulators make use of machine learning techniques like deep neural networks to learn the mapping from input parameters and/or intermediate representations such as dark matter fields to target observables, enabling the generation of realistic simulations at a fraction of the computational cost. In recent years, emulators have become a powerful shortcut in producing high-fidelity simulation of cosmological and astrophysical fields (see e.g.~\citep{perraudin2019cosmologicalnbodysimulationschallenge,Troster19,Li_2021,Harrington_2022,Horowitz_2022,mudur2023cosmological,jamieson_field-level_2023,Breitman_2023,Rigo_2025,hassan_hiflow_2022,andrianomena_emulating_2022,sharma25,bernardini2025ember2,andrianomena2024cosmologicalmultifieldemulator,mishra25,Riveros_2025,horowitz2025baryonbridge,mishra2026,Lavanderos26}), aiming at bridging the gap between dark matter simulations and baryonic observables. 

Early work~\cite{Troster19} used deep generative models such as GANs and VAEs to map dark matter fields to baryonic quantities in the context of the thermal Sunyaev–Zel’dovich effect, reproducing summary statistics but not addressing field-level inference. Later studies~\cite{Horowitz_2022, Harrington_2022} applied conditional VAEs and U-Nets to reconstruct hydrodynamical fields from N-body simulations, though only for fixed cosmological and astrophysical parameters, and a high redshift, relevant for the Lyman-$\alpha$ forest signal. More recent emulators~\cite{bernardini2025ember2} extended these approaches across cosmic time and multiple baryonic fields, using more sophisticated CNN architectures, aiming at lower-dimensional projected fields, although still without generalizing over the physical parameters space. Recent work~\cite{horowitz2025baryonbridge} has introduced conditional generative models that map fast particle-mesh simulations to full 3D baryonic fields while conditioning on both cosmological and astrophysical parameters, demonstrating accurate recovery of high redshift Ly$\alpha$-relevant statistics. Further work has been carried out in the development of generative diffusion models~\cite{PhysRevD.109.123536}; in this context, \cite{mishra25} introduces an algorithm to generate arbitrarily large 21 cm intensity fields, conditional on dark matter halo fields. In a parallel direction, some authors have also recently investigated physics-inspired particle-level approaches~\cite{LDL, mauro} to generate fast and accurate dark matter and baryonic fields using a small set of model parameters.

The authors of~\cite{Bayer_2025} highlight a critical challenge for SBI, namely its sensitivity to distributional shifts between different simulation codes using the same parameters and initial conditions (even just in the pure N-body regime). In particular, discrepancies arising from numerical schemes of cosmological N-body simulations and effective resolution can lead to out-of-distribution (OOD) behaviors and hence biased parameter inference. These effects appear to be driven by small-scale differences, to which neural networks are especially sensitive, and can sometimes be mitigated by applying appropriate smoothing to restore statistical consistency between datasets, although this comes at the expense of constraining power.

Finally, one last key aspect of modern LSS analyses relevant to this work is the joint exploitation of multiple tracers of the underlying matter distribution. Different observables probe the same large-scale structures while being affected by distinct astrophysical processes and observational systematics. As a result, their combination through cross-correlations provides a promising path to enhance the overall constraining power by reducing such systematics~\cite{Bonaldi:sys, Camera:sys}. Previous work includes the cross-correlation of different LSS tracers~\cite{Martinez:cross,Jain:cross,Yang:cross,Paech:cross}, LSS with the Cosmic Microwave Background~\cite{nolta:2004, ho:correlation, hirata:correlation, raccanelli:crosscorrelation, raccanelli:radio, raccanelli:isw, Bianchini:2014dla, Bianchini:2015fiw, Bianchini:2015yly, Mukherjee:gwxcmb}, the IM of different lines~\cite{Schmidt13:cross,Alonso15,Kovetz16:cross,Alonso16:cross,Wolz2016:cross,Pourtsidou16:cross,Pourtsidou16:cross_2,Raccanelli16:cross,Pourtsidou17:IM,Wolz17:cross,Alonso17:lssxim,Wolz18:cross,Cunnington18:lssxim}, neutrinos~\cite{fang:cross}, Gravitational Waves~\cite{Oguri:2016, raccanelli:pbhprogenitors, Scelfo18:gwxlss,Scelfo20:gws,namikawa:cross_ng,alonso:cross,Canas:sgwb,Calore:crosscorrelating,camera:gwlensing, Libanore+21, Mukherjee:gwxlss1, Mukherjee:gwxlss2,Mukherjee:sgwb,canas2021gaus, Cigarran22, Scelfo:gwxim, Scelfo_2023, Bosi_2023, semenzato2024cross}, or machine learning based multi-field analysis directly from simulations~\cite{villaescusa2021multifield}.

The goal of this work is to develop a proof-of-concept exploration of an SBI approach to multi-field LSS analysis, by developing a blueprint for a scalable SBI pipeline capable of inferring marginal posterior distributions for cosmological parameters, namely $\Omega_{\rm m}$ and $\sigma_8$. We focus on two complementary observational probes: galaxy number density fields and the mean brightness temperature of neutral hydrogen. By analyzing these tracers either individually or in combination, we aim to assess the gain in information when moving from single-tracer to joint analyses. We also investigate the information gain that can be expected from overcoming standard compression schemes like the power spectrum to machine-learning derived representations of full-field 3D inference. Emulators play a key role in this work, since they allow us to efficiently generate large numbers of galaxy and HI maps starting from dark matter-only simulations, thus bypassing the need for large numbers of hydrodynamical runs.
Our approach thus combines (i) fast approximate dark matter simulations, (ii) emulators to generate galaxy and HI fields, and (iii) SBI techniques to perform inference either on summary statistics or on latent representation of 3D fields. This allows us to systematically investigate the trade-offs between computational cost, data compression, and information retention, and to quantify gains and challenges of moving towards full-field 3D inference and multi-tracers analyses.

The manuscript is organized as follows. In section \ref{sec:methodology} we present the methodology, breaking it down in an overall overview (section \ref{sec:sketch}), the generation of dark matter simulations (section \ref{sec:DM_maps}) and the SBI framework (\ref{sec:SBI}). In section \ref{sec:results} we describe our results (on validation simulations) for SBI inference from the power spectrum alone and from the full 3D field, either for one or a combination of tracers. We draw our conclusions in section \ref{sec:conclusions}.

\section{Methodology}
\label{sec:methodology}

This section outlines the methodology of this pipeline, aiming to develop a flexible SBI pipeline capable of extracting cosmological information directly from large-scale structure fields and assess the advantages and limitations of field-level SBI on multiple tracers. 

We adopt a controlled and idealized setup that allows us to isolate the cosmological information content of the simulated fields. In particular, we account for cosmic variance by considering multiple realizations and using an averaging procedure over initial conditions, thus ensuring that inference is not driven by specific initial conditions. At the same time, we deliberately do not include additional sources of observational uncertainty---such as instrumental noise, survey systematics, or foreground contamination---in order to establish a clear and robust baseline of an idealized scenario for the performance of the pipeline. Our training, validation and testing sets are all self-consistently based on emulated maps, given that thoroughly testing the validity of transfer learning to the original suite (or to different ones) would require a full OOD analysis that go beyond the scope of this idealized setup work. In this sense, our analysis should be regarded as a proof-of-concept, designed to demonstrate the feasibility, potential and limitations of the proposed SBI framework, especially in the field-level and multi-tracers approaches. Building on these results, a natural next step will be to incorporate increasingly realistic observational effects and to tailor the analysis to the characteristics of ongoing and future surveys, further enhancing the applicability of our methodology.

\subsection{Overview of pipeline}
\label{sec:sketch}

\begin{figure}[tb]
    \centering

    \begin{minipage}{0.9\textwidth}
      \centering
      \includegraphics[width=\linewidth]{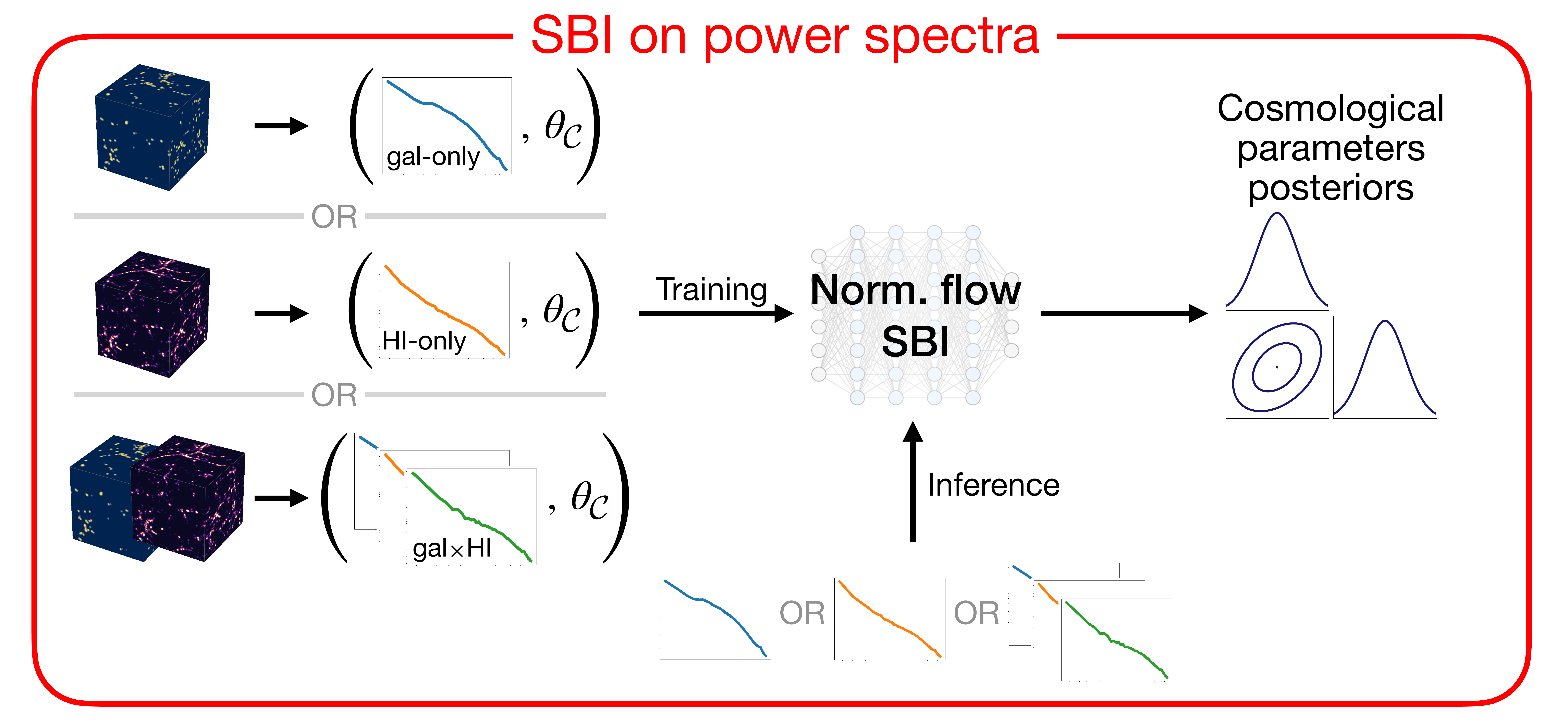}
    \end{minipage}

    \vspace{0.5cm}

    \begin{minipage}{1.00\textwidth}
      \centering
      \includegraphics[width=\linewidth]{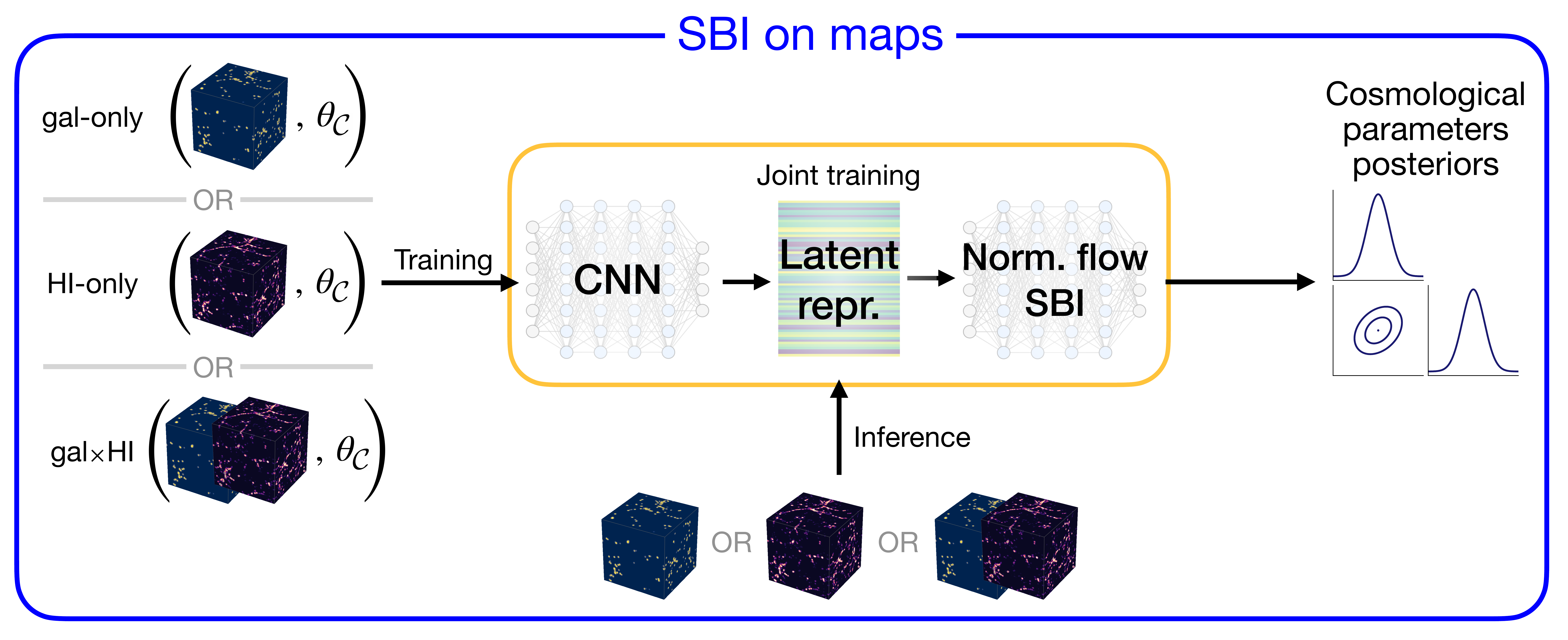}
    \end{minipage}

    \caption{\textit{Top}: sketch of the SBI on power spectra pipeline. Power spectra are extracted from the emulated galaxy and/or HI maps, so that pairs of power spectra and cosmological parameters $\theta_{\mathcal{C}}=\{\Omega_{\rm m}, \sigma_8\}$ constitute the SBI training set (observables fields are reduced to power spectra). Once trained, the SBI flow is able to produce posteriors on $\theta_{\mathcal{C}}$ for validation power spectra. \textit{Bottom}: sketch for SBI on maps pipeline. Pairs of emulated full maps and $\theta_{\mathcal{C}}$ are passed for joint training to a Convolutional Neural Network (CNN) which extracts a latent representation of the data, which is then used to train the SBI normalizing flow. 
    }
    \label{fig:pipelines_sketch}
\end{figure}

Before detailing each component of our method, here we provide a high-level description:
\begin{enumerate}
	\item In order to obtain observable predictions for LSS, we start from approximate dark matter (DM)-only 3D maps, evolved from high redshift down to $z=0$ with \texttt{FastPM}~\cite{FastPM}. Maps feature a box size of either 25 or 50$\,h^{-1}$Mpc; additional details are given in section \ref{sec:DM_maps}.
	\item  We train a modified version of the the U-Net~\cite{ronneberger2015unetconvolutionalnetworksbiomedical} emulator presented in~\cite{mishra25}, \texttt{HALOgen}, to generate, conditioned on the DM maps, 3D maps of two different observables: galaxy number counts density and mean brightness temperature $T_b$ of neutral hydrogen. We refer to such emulator as \hgl{} (\texttt{HALOgen}-like). For emulator training, this work takes advantage of the hydrodynamical simulations from the Latin Hypercube (LH) suite of the \texttt{CAMELS} Simulation Project~\cite{camels} with prescriptions from the \texttt{IllustrisTNG}~\citep{nelson2021illustristngsimulationspublicdata,Pillepich_2017} suite. Such simulations are available for a different combination of cosmological parameters ($\Omega_{\rm m}$, $\sigma_8$) and astrophysical ones ({$\rm A_{SN1},A_{AGN1},A_{SN2},A_{AGN2}$}), describing the relative strength and efficiency of feedback processes from Supernovae and Active Galactic Nuclei. More details on observables construction are provided in section \ref{sec:observables}, while for details on the emulator training and architecture  we refer the reader to appendix \ref{app:emulator}.
	\item We perform SBI on emulated galaxy/HI maps to infer (marginal) posterior distributions on cosmological parameters. We investigate two approaches: first, we perform SBI on a summary statistic of the full field, the power spectrum $P(k)$ (see top panel of figure \ref{fig:pipelines_sketch}); second, we perform SBI on a machine learning representation of the full field (see bottom panel of figure \ref{fig:pipelines_sketch}). Within each, we compare the constraints on cosmology that can be achieved from one single tracer (galaxy or HI) or from simultaneously considering both (different input lines in figure \ref{fig:pipelines_sketch}, bottom).
    See section \ref{sec:SBI} for full details. 
\end{enumerate}

Figure \ref{fig:pipelines_sketch} provides a schematic overview of our pipeline, with each step described in detail below. In this work, all observables are considered at redshift $z$=0, working in the plane-parallel approximation and assuming
the box redshift to be constant. Extending the analysis to higher $z$ is planned for future works.

\subsection{Dark matter approximate maps}\label{sec:DM_maps}
We train the \hgl{} emulator to produce 3D galaxy/HI fields starting from an approximate, $z=0$ 3D DM map. For the purpose of constructing a suitable training set for the emulator, we make use of \texttt{FastPM} to generate approximate DM maps at small computational cost, and we pair them with galaxy/HI 3D maps obtained from full hydro simulations from \texttt{CAMELS} for the same cosmological parameters and initial conditions (further details on galaxy emulator train/validation sets in appendix \ref{app:emulator}). \texttt{FastPM} takes as inputs the cosmological parameters $\{\Omega_{\rm m}, \sigma_8\}$ (the latter in the form of the normalization of the input power spectrum), the linear total matter power spectrum, and a seed for the generation of the linear density map at high redshift. 

The \texttt{FastPM} random number generator is compatible with the one used for \texttt{CAMELS} (part of \texttt{N-G{\footnotesize EN}IC}~\cite{ics}), therefore, for the training set, we simply feed \texttt{FastPM} the seeds provided in the \texttt{CAMELS} dataset, as well as the linear power spectrum also provided in the \texttt{CAMELS} simulation suite data. On the contrary, when emulating observables from new DM maps (i.e. when training and testing the SBI model), we generate the linear power spectrum using the Boltzmann solver \texttt{CLASS}~\cite{class} and we feed \texttt{FastPM} arbitrary seeds.

We generate DM cubes with both $L=25\,h^{-1}$Mpc  and $L=50\,h^{-1}$Mpc sides (we discuss this choice more in detail below), with the following properties:
\begin{itemize}
	\item $L=25\,h^{-1}$Mpc: \texttt{FastPM} is run with $128^{3}$ particles on a $128^{3}$ mesh and maps are saved with a mesh size of $64^{3}$;
	\item $L=50\,h^{-1}$Mpc: \texttt{FastPM} is run with $256^{3}$ particles on a $256^{3}$ mesh and maps are saved with a mesh size of $128^{3}$.
\end{itemize}
With these specifications, maps of different sizes have the same resolution. In each case, we generate a linear density map with twice the resolution of the mesh used for the simulator in order to improve the accuracy on small scales. We then generate initial conditions at redshift $z=9$ using \texttt{2LPT}~\cite{2lpt} and run 10 \texttt{FastPM} steps down to $z=0$ using a force resolution $B=2$, with an overall compute time of $\sim 0.3$ CPU hours per simulation ($\sim35$ seconds on one node of the Leonardo cluster of CINECA).

\subsection{Galaxies and HI maps}\label{sec:observables}
The first tracer we consider consists of galaxy number counts density maps. To generate these maps, we carry out the following steps: we choose the full hydrodynamical snapshots from the \texttt{CAMELS} LH suite with comoving volume $L^3 = {(25 \:h^{-1}{\rm Mpc}})^{3}$ and $N_\mathrm{part} = 256^3$ DM particles and gas elements, at $z=0$. We apply a mask selecting sub-halos with mass larger than a threshold\footnote{This threshold is chosen for two reasons: first, we would like to have numerically resolved haloes and this lower limit guarantees at least 100 dark matter particles per halo; second, we would like to use all the haloes which are likely to host HI according to more sophisticated hydro sims like \cite{villaescusa-navarro_ingredients_2018}.} of $10^9 M_{\odot}/h$ and non-zero stellar mass, and generate the density field with the cloud-in-cell (CIC) mass-assignment scheme on a grid of size $N_{\rm grid} = 64$, further applying a smoothing with a Gaussian kernel of radius $R = 0.2\,h^{-1}$Mpc (following \cite{mishra25}) and, to help training, clipping low values of the density field to a minimum of $10^{-4}$. 

Maps of the second tracer, HI temperature, are obtained from the correspoding full hydrodynamic snapshots with the same $L^3$ and $N_\mathrm{part}$ values, at $z=0$. We extract the neutral hydrogen fraction $x_{\mathrm{HI}, i}$ and gas mass $m_{g,i}~[M_\odot]$ for each gas particle $i$ in the simulation. The corresponding neutral hydrogen density is calculated as 
\begin{equation}
\rho_{\text{HI}, i} = m_{g,i} X x_{\text{HI}, i} \left(\frac{N_\text{part}}{a\times L}\right)^3 \left[\frac{\text{kg}}{\text{m}^3}\right]\, , 
\end{equation}
where $X=0.76$ is the hydrogen mass fraction as set in \texttt{CAMELS} and $a$ is the scale factor. We set $x_{\mathrm{HI}, i} = 1$ for particles with non-zero star formation rate, following the prescription adopted in \texttt{CAMELS}~\cite{villaescusa-navarro_camels_2022}. This approximation assumes complete self-shielding and reproduces key observational constraints, including the HI abundance and the column density distribution of damped Lyman-$\alpha$ systems~\cite{villaescusa-navarro_ingredients_2018}. We then interpolate $\rho_{\mathrm{HI}, i}$ onto a continuous field $\rho_{\mathrm{HI}}$ using the CIC scheme and $N_{\rm grid} = 64$. The 21$\,$cm brightness temperature in the post-reionization regime is approximated as~\cite{villaescusa-navarro_ingredients_2018} 
\begin{equation}
T_b(\textbf{x}) = 189h\left(\frac{H_0(1+z)^2}{H(z)}\right)\frac{\rho_{\text{HI}}(\textbf{x})}{\rho_c} ~[\text{mK}]\, ,
\end{equation}
where $\rho_{\mathrm{HI}}(\mathbf{x})$
is the interpolated neutral hydrogen density at position $\mathbf{x}$, $H(z)$ is the Hubble--Lemaître parameter, $h=0.6711$ is the dimensionless Hubble--Lemaître constant (fixed at \texttt{CAMELS} fiducial value) and $\rho_c=9.47\cdot 10^{-27}kg/m^3$ is the critical density. This field is also smoothed with a $R = 0.2\,h^{-1}$Mpc Gaussian kernel.

The process of predicting such 3D galaxy and HI fields from their approximate DM counterpart can be thought of an image to image translation task. To achieve this while preserving spatial correlations and structure we make use of \hgl. While the original \texttt{HALOgen} \cite{mishra25} was designed to work for fixed cosmological parameters, here we introduce structural modifications to generalize emulated fields for different cosmological $\cosmo =\{\Omega_{\rm m}, \sigma_8 \}$ and astrophysical $\astro =\{ \rm A_{SN1},A_{AGN1},A_{SN2},A_{AGN2}\}$ parameters (defined as in \texttt{CAMELS}). 

One limitation of using relatively small boxes is the power spectrum sensitivity to cosmic variance and finite-volume effects. Indeed, we found that, for $25\: h^{-1}{\rm Mpc}$ boxes, realizations at fixed physical parameters and generated with different random seeds can yield noticeably different power spectra (creating a degeneracy between cosmology and the particular realization of the density field); for this reason, we adopt boxes of size $50\: h^{-1}{\rm Mpc}$ to mitigate this effect and improve the sampling of large-scale modes. However, training \hgl{} for 50$\,h^{-1}$Mpc boxes is not feasible for the purpose of this work within the \texttt{CAMELS} simulations, since the LH set is not available for a box size of 50$\,h^{-1}$Mpc, and the available SB35 set for such box size varies 30 astrophysical parameters at the same time (instead of 4), making the emulation task significantly more difficult. For these reasons, we train \hgl{} using the LH simulations at 25$\,h^{-1}$Mpc, and then use it to emulate maps at 50$\,h^{-1}$Mpc. Note that emulating maps of different size is correct only wheen keeping the same resolution, motivating the choice of number of particles and grid size for the \texttt{FastPM} simulations mentioned above. Details about the emulator architecture and training are provided in appendix \ref{app:emulator}.

\subsection{Simulation-Based Inference with Neural Posterior Estimation}\label{sec:SBI}
We use the galaxy and HI emulated maps to perform SBI in order to obtain posterior distributions of the cosmological parameters $\cosmo$; this is done by either fixing the 4 astrophysical parameters $\astro$ to fiducial values \{$\rm A_{\rm SN1}=A_{\rm AGN1}=A_{\rm SN2}=A_{\rm AGN2}=1$\} or by marginalizing over them. SBI is trained on newly generated maps, in terms of both seeds and parameters values, obtained via the emulator from uniform priors as follows: $\Omega_{\rm m} \in [0.2, 0.4]$ and $\sigma_8 \in [0.7,0.9]$ and, for the free astrophysics scenario, $\rm A_{SN1}\in [0.7, 1.5], A_{AGN1}\in [0.7, 1.5], A_{SN2}\in [0.5, 2.0], A_{AGN2}\in [0.5, 2.0]$.

We perform SBI with Neural Posterior Estimation (NPE)~\cite{Papamakarios2016arXiv160506376P,Lueckmann2017arXiv171101861L,Greenberg2019arXiv190507488G,Deistler2022arXiv221004815D}, in which a neural density estimator is trained to approximate the conditional posterior distribution $p(\params \mid \data)$ directly from simulated parameters-data pairs, $(\params, \data)$. In this work, the posterior is modeled using a normalizing flow~\cite{Kobyzev:NF}, i.e. an expressive, invertible transformation that maps a simple base distribution (typically a multivariate Gaussian) into a complex target distribution through a sequence of learned bijective transformations. NPE training maximizes the log-probability of sampled parameters given simulated data, i.e. minimizing the negative log-likelihood $\mathcal{L}(\phi) = -\mathbb{E}_{p(\params, \data)}[\log q_\phi(\params \mid \data)]$, where $q_\phi(\params \mid \data)$ denotes a parametric approximation of the true posterior and $\phi$ are the normalizing flow parameters (see e.g.~\cite{Papamakarios2016arXiv160506376P, Greenberg2019arXiv190507488G} for further details).

Specifically, we adopt a Masked Autoregressive Flow (MAF)~\cite{papamakarios2017masked}, a type of normalizing flow that uses autoregressive neural networks to transform a simple probability distribution into a complex one by masking weights to ensure each dimension only depends on previous ones. The main advantage of MAFs is that they provide highly efficient density estimation, allowing for relatively quick training. 

As anticipated in section \ref{sec:sketch}, we perform SBI on two different types of data:
\begin{itemize}
	\item {\bf SBI on power spectrum} (case $\mathcal{P}$): we perform SBI on the power spectra $P(k)$ extracted from the emulated galaxy/HI maps, as a baseline case of the achievable constraints within an SBI framework with this summary statistic alone. Firstly, we perform SBI looking only at one tracer (either galaxy or HI); then we take into account the full cross-correlation g$\times$HI information, given by both the auto-correlation power spectra $P_k^{\rm g}$ and $P_k^{\rm HI}$ and the cross-correlation power spectrum $P_k^{\rm g\times HI}$. For the sake of simplifying the notation, in what follows these three cases are referred to as, respectively, $\mathcal{P}^{\rm g}$, $\mathcal{P}^{\rm HI}$ and $\mathcal{P}^{\rm g \times HI}$. Full details on the methodology are provided in section \ref{sec:SBI-on-Pk}.
	\item {\bf SBI on latent representations of full maps} (case $\mathcal{M}$): the entire density field is used here, to inform a latent representation of the full galaxy/HI maps, extracted by a Convolutional Neural Network (CNN) attached to the SBI flow for joint training. As in the previous scenario, we first work on one single field and then feed both of them to the network jointly. Such cases are referred to as $\mathcal{M}^{\rm g}$, $\mathcal{M}^{\rm HI}$ and $\mathcal{M}^{\rm g \times HI}$. We explore both the case where maps are flattened (2D$\mathcal{M}$) by averaging along a third axis, and a full three-dimensional scenario (3D$\mathcal{M}$), which is more challenging from a memory and training perspective, while providing in principle more information. Full details are provided in section \ref{sec:SBI-on-maps}.
\end{itemize}
Table \ref{tab:sbi_options} details the different SBI setups explored and summarizes the nomenclature that identifies each configuration.

\begin{table}[t]
\centering
\caption{Summary of the different SBI setups explored in this work grouped by category, and corresponding abbreviations. For example: SBI on power spectrum performed with fixed astrophysics and galaxies as single tracer is referred to as $\rm \mathcal{P}^{\rm g}_{\rm fixedA}$ and SBI on maps performed on 3D cubes of both observables, with free astrophysics is referred to as $\rm 3D\mathcal{M}^{\rm g\times HI}$.}
\label{tab:sbi_options}
\renewcommand{\arraystretch}{1.3}
\resizebox{\columnwidth}{!}{\begin{tabular}{|c|c|}
\hline
\textbf{CATEGORY} & \textbf{OPTIONS} \\
\hline
Data representation & $P(k)$: $\mathcal{P}$ \quad representation of full-field maps: 2D$\mathcal{M}$ or 3D$\mathcal{M}$ \\
\hline
Astrophysics & fixed: subscript ``fixedA'' \quad free: no subscript \\
\hline
tracers & gal-only: apex ``g'' \quad HI-only: apex ``HI'' \quad galaxies$\times$HI: apex ``g$\times$HI'' \\
\hline
\end{tabular}}
\end{table}

The inference pipeline is implemented using the \texttt{sbi} library~\cite{sbi_library}. We test different hyper-parameters configurations; the chosen values are specified in each corresponding subsection.

\subsubsection{Case $\mathcal{P}$: SBI on Power Spectrum}\label{sec:SBI-on-Pk}
For the SBI analysis performed on power spectrum (case $\mathcal{P}$), we compress the three-dimensional information of the emulated galaxy/HI field into its isotropic power spectrum $P(k)^{\mathrm X}$, with $X \in \{\rm g, HI\}$. We compute it by means of the \texttt{Pylians} library~\cite{Pylians}: starting from the density field defined on a regular grid, we perform a Fast Fourier Transform to obtain the Fourier modes and estimate the spherically-averaged power spectrum, binning modes in shells of constant wavenumber $k$: $P(k)^{\mathrm X} = \langle |\delta^{\mathrm X}(k)|^2 \rangle$, where $\delta^{\mathrm X}(k)$ is the Fourier transform of the galaxy number density/HI mean brightness temperature field (both normalized by their mean). Similarly, for the cross-correlation case we compute $P(k)^{\mathrm X, Y} = \langle \delta^{\mathrm X}(k) \delta^{\mathrm Y^*}(k) \rangle$ with ${\rm X \neq Y}$. Computations are performed over the range of scales allowed by the box size and grid resolution.

We consider power spectra extracted from boxes of size $50\: h^{-1}{\rm Mpc}$, in the scales range $k \in [0.18, 4.7] \: h$/Mpc. For each combination of physical parameters we average the power spectra obtained from 10 independent realizations with different initial seeds. This procedure reduces the impact of realization-to-realization fluctuations and is effectively equivalent to probing a larger survey volume, as it increases the number of independent modes contributing to the estimated power spectrum. We remind the reader that in our simplified setup no measurement noise is added to the simulated data. In this regard, the power-spectrum-related uncertainties treatment includes only the aforementioned contribution from cosmic variance.

To build the SBI training set for the $\mathcal{P}$ case, we emulate $8 \times 10^4$ galaxy/HI maps drawing physical parameters from their prior ranges and producing 10 emulations (each with different initial conditions) for each parameter set value, over which we average the power spectra. This produces $8 \times 10^3$ averaged power spectra. The validation set is obtained as 20\% of the training set, randomly sampled (after averaging over initial conditions). The SBI test set is made of 200 averaged (each over 10 initial conditions) power spectra, all drawn from the fiducial $\Lambda$CDM scenario with $\cosmo = \{\Omega_{\rm m}=0.315,\: \sigma_8=0.811\}$ from Planck fiducial values~\cite{planck18} and either with $\astro = \{\rm A_{\rm SN1}=A_{\rm AGN1}=A_{\rm SN2}=A_{\rm AGN2}=1\}$ (for the fixed astrophysics baseline case) or marginalizing over different astrophysics values within their uniform priors. After the averaging and before passing the power spectra to the SBI flow, we move to $\log_{10}$ space and normalize the sets with standard scaling.

For the cross-correlation cases, $\rm \mathcal{P}^{\rm g\times HI}_{\rm fixedA}$ and $\rm \mathcal{P}^{\rm g\times HI}$, we pass to the SBI flow a concatenation of both the auto-spectra and the cross-spectrum, after they have been re-normalized all together: $\{P(k)^{\rm g}, P(k)^{\rm HI}, P(k)^{\rm g, HI}\}$.

The MAF network adopted to model the posterior distribution is configured with 5 transforms and 128 hidden features per layer. Training is performed with a batch size of 128 and a learning rate of $10^{-3}$. We train with early stopping applied if the validation loss does not improve for 120 consecutive epochs. Such hyper-parameters configuration are chosen after exploring several different combinations.

Section \ref{sec:results} provides results for posteriors on $\{\Omega_{\rm m}, \sigma_8\}$ and calibration plots both fixing the astrophysics to fiducial values and letting it free, marginalizing over the four astrophysical parameters.

\subsubsection{Case $\mathcal{M}$: SBI on latent representations from maps}\label{sec:SBI-on-maps}
Field-level SBI relies on a CNN to compress the input emulated maps into a one dimensional latent space, which is then passed to the normalizing flow. The CNN encoder and the flow are trained jointly in order to obtain a physically informative latent space. Compressing the maps by means of a separate autoencoder guided by a reconstruction loss would not necessarily preserve the most informative features for the physical parameters of interest.

In a computationally less demanding configuration, we flatten the full maps to 2 dimensions by averaging over the third axis, and extract the latent space from them. We refer to this scenario as 2D$\mathcal{M}$ case. We also explore the increase of information provided by the entire three-dimensional map. This scenario, referred to as 3D$\mathcal{M}$ case, is ideally expected to provide the highest amount of information, but at a significantly higher computational cost.

The SBI train set is here composed of $8 \times 10^4$ ($3 \times 10^4$) maps of box size 50$\,h^{-1}$Mpc and $N_{\rm grid}=128$ for the 2D (3D) case, the 3D set being smaller for computational costs reasons. To account for cosmic variance and finite box effects, 10 different initial conditions are adopted for each parameter value, as was done in the $\mathcal{P}$ case. The test set is instead made of 200 $\Lambda$CDM-Planck realizations, with different never-seen-before seeds, same cosmology $\cosmo = \{\Omega_{\rm m}=0.315,\: \sigma_8=0.811\}$~\cite{planck18} and either with fixed astrophysics ($\astro = \{\rm A_{\rm SN1}=A_{\rm AGN1}=A_{\rm SN2}=A_{\rm AGN2}=1\}$) or marginalizing over astrophysical effects within their priors.

The CNN encoder is implemented in either 2D or 3D depending on the dimensionality of the input maps, with the same architecture applied in both cases. It is composed of two layers with 32 and 64 filters, respectively, using kernel size 3 and stride 2 in each layer and a ReLU activation function. We adopt group normalization, with 4 and 8 groups in the first and second layers, respectively. In the single tracer cases, $\mathcal{M}^{\rm g}$ and $\mathcal{M}^{\rm HI}$, the network is characterized by only one channel, whereas in the cross-correlation case, $\mathcal{M}^{\rm g \times HI}$, we have one channel for each tracer, for a total of two. The final latent space is obtained via average pooling over the spatial dimensions and has a single dimension of length 64 for all cases.

Similarly to the $\mathcal{P}$ case, we model the posterior using a MAF with 5 transforms and 128 hidden features per layer. The network is trained with a learning rate of $10^{-3}$, a batch size of 128 (16) with early stopping if the validation loss does not improve for 120 (35) epochs for the 2D (3D) case. We reserve 20\% of the simulations for validation. Due to the extreme computational costs of full-field SBI, especially for the 3D case, we could not perform parameters tuning as extensively as in the $\mathcal{P}$ case. We chose, instead, configurations analogue to the best $\mathcal{P}$ case ones. Table \ref{tab:sbi_architecture} summarizes the architecture  for the $\mathcal{M}$ cases.

To generate the results below, the $\mathcal{P}$ cases were trained for around $2$ minutes on a Apple McBookPro M4 CPU, the 2D$\mathcal{M}$ took about 40 minutes on a single A100 GPU whereas training the 3D$\mathcal{M}$ cases required one full week on the same single A100 GPU (although preliminary posteriors near convergence were already available after a runtime of $\sim2$ days). In all cases, inference requires only a few seconds on the Apple McBookPro M4 CPU. In terms of disc usage for the 50$\,h^{-1}$Mpc case, the full train set for the $\mathcal{P}^{\rm g \times HI}$ scenario occupies a mere $\sim 1\,$MB ($8 \times 10^3 \times 3$ average power spectra), the one for the 2D$\mathcal{M}^{\rm g \times HI}$ case $\sim 10\,$GB ($8 \times 10^4 \times 2$ maps) and finally $\sim 470\,$GB for 3D$\mathcal{M}^{\rm g \times HI}$ ($3 \times 10^4 \times 2$ cubes). 

\begin{table}
\centering
\caption{Architecture of the SBI model for 3D field-level inference (model 3D$\mathcal{M}$). The input maps are compressed into a 64-dimensional latent representation via a convolutional encoder, and the posterior over parameters is inferred using NPE with a MAF. Note that $d=2$ (3) for 2D (3D) field-level inference.}
\label{tab:sbi_architecture}
\begin{tabular}{llll}
\hline
\textbf{Stage} & \textbf{Representation} & \textbf{Channels} & \textbf{Operation} \\
\hline
Input & $128^d$ maps & 1 or 2 & Input fields \\
Encoder  & $(128/2)^d$ & 32 & Conv (k=3, s=2) + GN (g=4) + ReLU\\
 { } & $(128/4)^d$ & 64 & Conv (k=3, s=2) + GN (g=8) + ReLU \\
Latent & $\mathbb{R}^{64}$ & -- & Global mean pooling \\
Flow & $\mathbb{R}^{64}$ & -- & MAF (5 transforms, 128 hidden features) \\
Output & $p( \params \mid \data)$ & -- & Approximate posterior distribution \\
\hline
\end{tabular}
\end{table}

\section{Results}
\label{sec:results}

\subsection{Statistical precision}
\begin{figure}[h!]
	\centering
	\includegraphics[width=\linewidth]{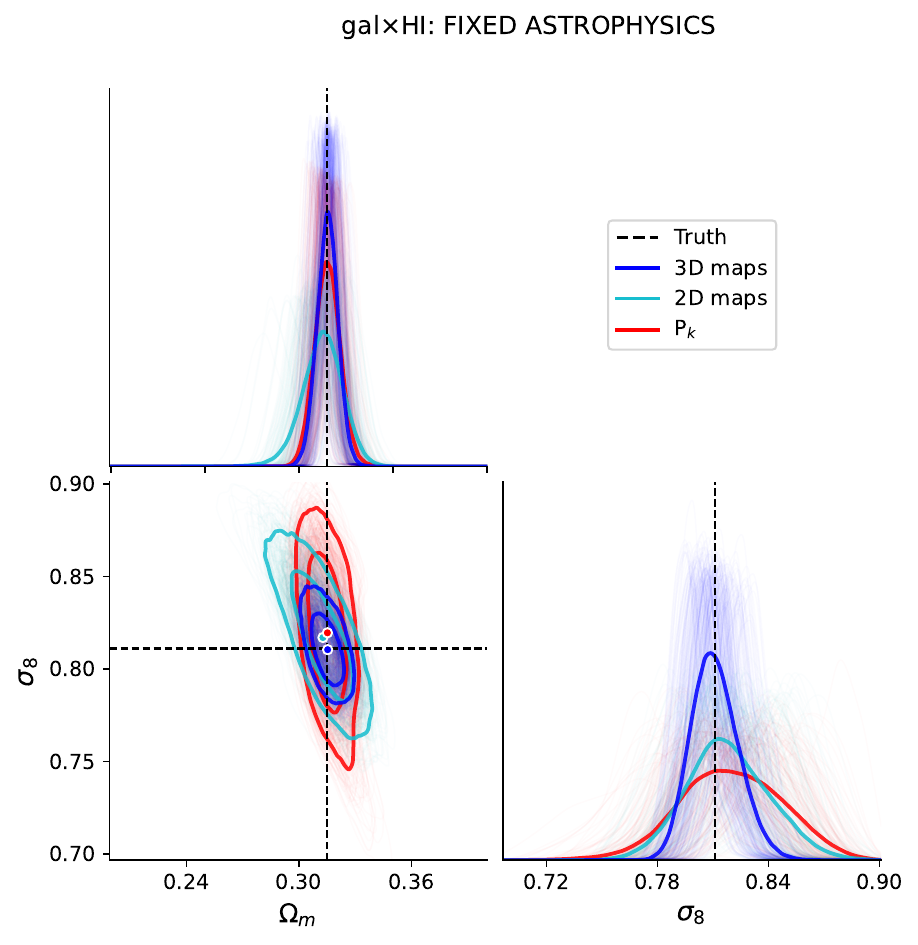}
	\caption{Posteriors on $\{\Omega_{\rm m}, \sigma_8\}$ for $\mathcal{P}^{\rm g \times HI}_{\rm fixedA}$ (red) vs $\mathcal{M}^{\rm g \times HI}_{\rm fixedA}$ (2D, light blue, and 3D, blue) for 200 $\Lambda$CDM-Planck realizations test set (ground truth in black). Performing inference on the full 3D maps significantly improves precision. Dots in the lower left plot represent the median values computed from all the posteriors samples.}
    \label{fig:posteriors_SBI-on-Pkvsmap_fixedastro}
\end{figure}

\begin{figure}[h!]
	\centering
	\includegraphics[width=\linewidth]{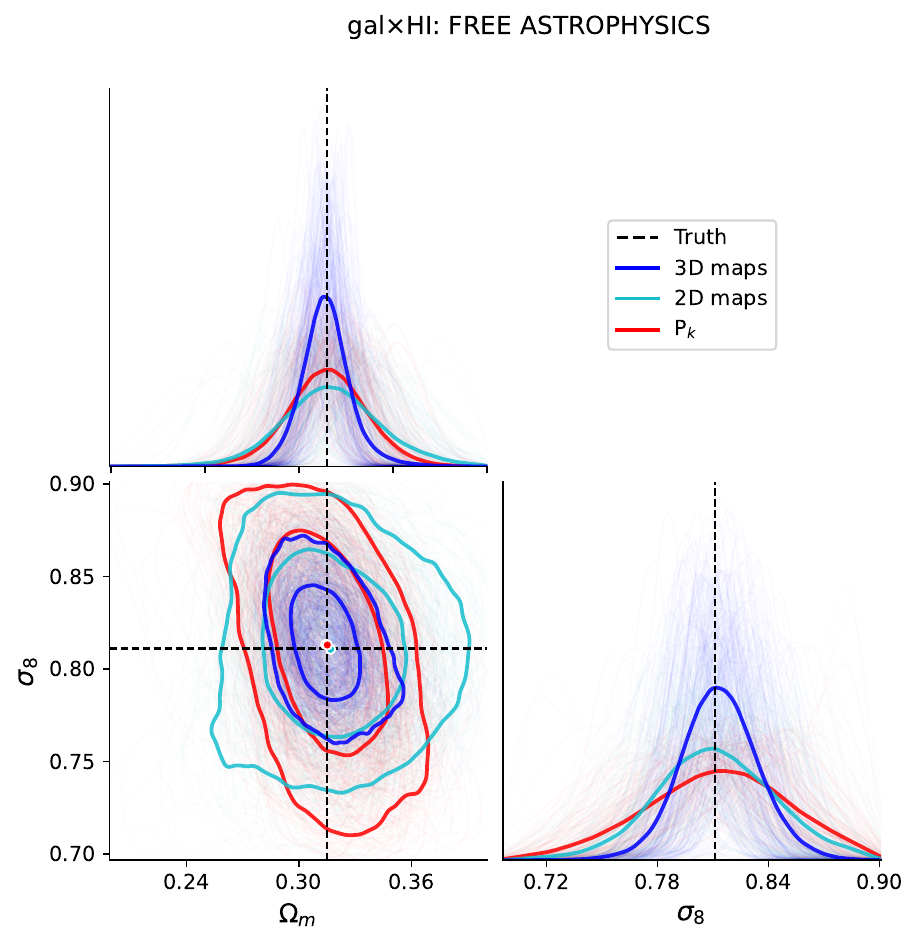}
	\caption{As in figure \ref{fig:posteriors_SBI-on-Pkvsmap_fixedastro}, but for the free astrophysics scenario, with astrophysical parameters marginalized over.}
    \label{fig:posteriors_SBI-on-Pkvsmap_freeastro}
\end{figure}

\begin{figure}[h]
	\centering
	\includegraphics[width=\linewidth]{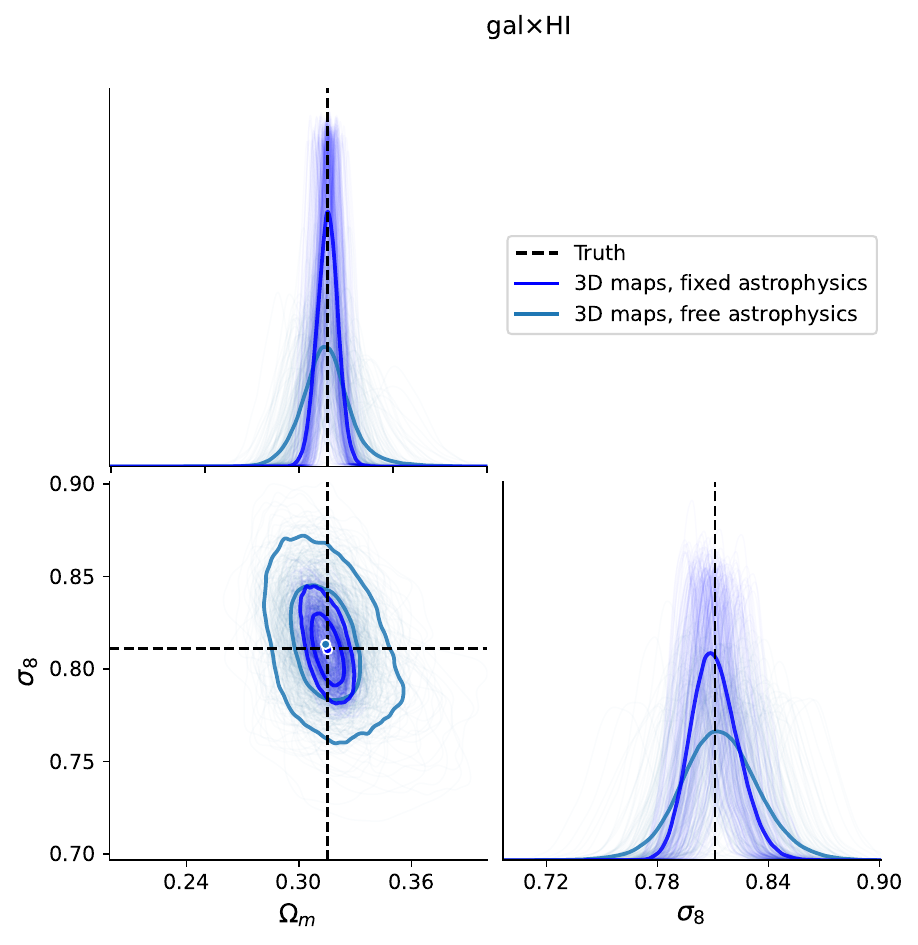}
	\caption{Posteriors on $\{\Omega_{\rm m}, \sigma_8\}$ marginalizing over the free astrophysics case 3D$\mathcal{M}^{\rm g \times HI}$ versus fixed astrophysics 3D$\mathcal{M}^{\rm g \times HI}_{\rm fixedA}$ for 200 validation datasets (ground truth in black). Fixing astrophysics shows smaller uncertainties as expected, but marginalized constraints retain cosmological information while being agnostic to the details of baryonic processes.}
    \label{fig:posteriors_SBI-on-maps_astrocases}
\end{figure}

\begin{figure}[h]
	\centering
	\includegraphics[width=\linewidth]{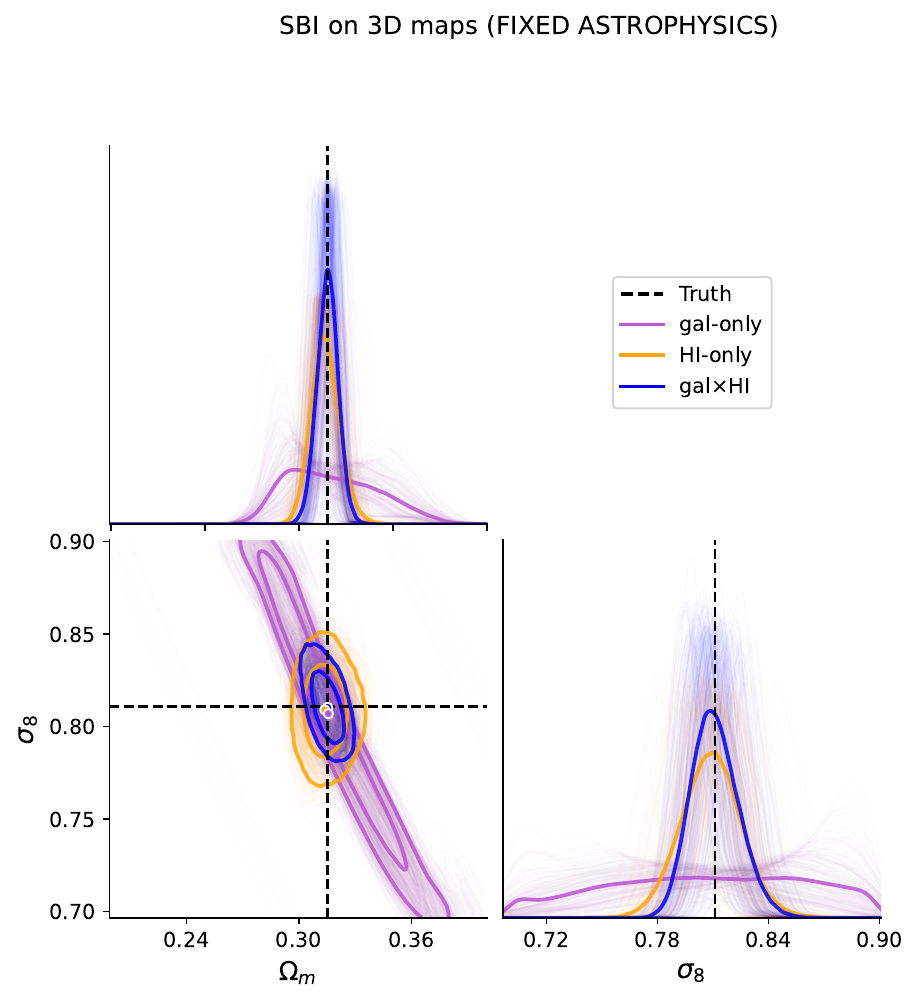}
	\caption{Posteriors on $\{\Omega_{\rm m}, \sigma_8\}$ (fixed astrophysics) comparing single-tracers cases 3D$\mathcal{M}^{\rm g}_{\rm fixedA}$ (purple) and 3D$\mathcal{M}^{\rm HI}_{\rm fixedA}$ (orange) with multi-tracer case 3D$\mathcal{M}^{\rm g \times HI}_{\rm fixedA}$ (blue).}
    \label{fig:posteriors_singleVSmulti_fixedastro}
\end{figure}

\begin{figure}[h]
	\centering
	\includegraphics[width=\linewidth]{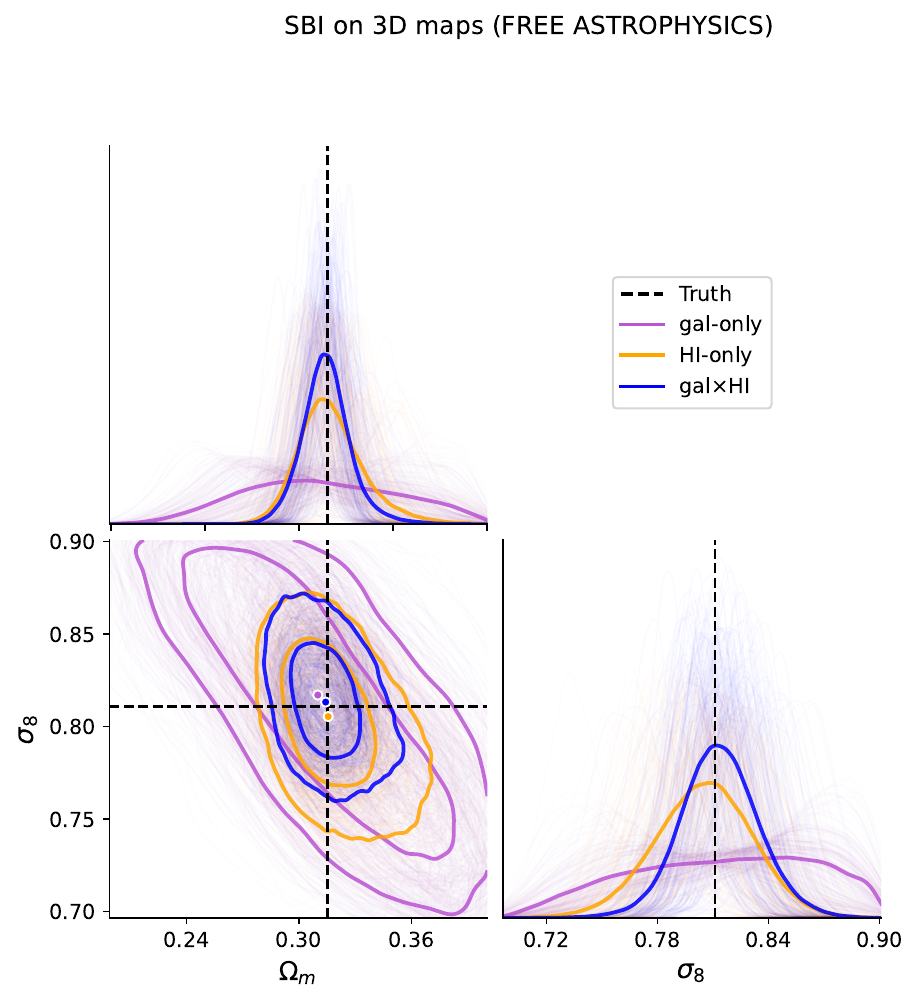}
	\caption{As in figure \ref{fig:posteriors_singleVSmulti_fixedastro}, but for the free astrophysics scenario.}
    \label{fig:posteriors_singleVSmulti_freeastro}
\end{figure}

Figure \ref{fig:posteriors_SBI-on-Pkvsmap_fixedastro} shows posteriors on $\{\Omega_{\rm m}, \sigma_8 \}$ for the 200 realizations in the test set (at fixed $\Lambda$CDM Planck ground truth parameters) for the baseline fixed-astrophysics scenario, giving an overview of the gains reachable with field-level SBI. The following cases are shown: $\mathcal{P}^{\rm g \times HI}_{\rm fixedA}$, 2D$\mathcal{M}^{\rm g \times HI}_{\rm fixedA}$ and 3D$\mathcal{M}^{\rm g \times HI}_{\rm fixedA}$. The light lines indicate contours for each test set realization, whereas solid lines provide the mean posterior, averaged overall all realizations. Although the 2D$\mathcal{M}^{\rm g \times HI}_{\rm fixedA}$ case does not significantly improve constraining power, the 3D$\mathcal{M}^{\rm g \times HI}_{\rm fixedA}$ case provides a significant gain, especially for $\sigma_8$. An even clearer improvement can be seen when marginalizing over the astrophysical parameters: figure \ref{fig:posteriors_SBI-on-Pkvsmap_freeastro} displays significant improvements on both $\{\Omega_{\rm m}, \sigma_8 \}$ when moving to 3D field-level (case 3D$\mathcal{M}^{\rm g \times HI}$). In figure \ref{fig:posteriors_SBI-on-maps_astrocases} we display posteriors for the free astrophysics 3D$\mathcal{M}^{\rm g \times HI}$ case, comparing it with the baseline 3D$\mathcal{M}^{\rm g \times HI}_{\rm fixedA}$ scenario. Although marginalization over astrophysical parameters reduces the precision of marginal posteriors by about a factor $\sim 2$ for both $\Omega_{\rm m}$ and $\sigma_8$, additional cosmological information is still clearly retained in the field-level approach. This is different for the $\mathcal{P}^{\rm g \times HI}$ case previously shown in figure \ref{fig:posteriors_SBI-on-Pkvsmap_freeastro}, where it can be noticed that limiting the analysis to the power spectrum provides very wide posteriors, especially on $\sigma_8$ where the posterior spans nearly the whole prior range. Including information beyond the power spectrum seems therefore crucial to retain cosmological information while marginalizing over astrophysics. Figure \ref{fig:posteriors_singleVSmulti_fixedastro} (fixed astrophysics) and figure \ref{fig:posteriors_singleVSmulti_freeastro} (free astrophysics) also help highlighting the improvement obtainable in terms of posteriors predictivity when moving from single tracer to multi-tracers analysis. In particular, we notice how including HI information produces a dramatic improvement in constraints compared to galaxy tracers only. 

\begin{figure}[htbp]
\centering

\begin{minipage}{0.48\textwidth}
\centering
\renewcommand{\arraystretch}{1.4}
\small
\begin{tabular}{@{}cccc@{}}
\toprule
\makecell{\textbf{Data}\\\textbf{summary}} &
\textbf{Tracer} &
\makecell{\textbf{FoM$_m$}\\\textbf{(Fixed A.)}} &
\makecell{\textbf{FoM$_m$}\\\textbf{(Free A.)}} \\
\midrule
$\mathcal{P}$ & g & $0.65$ & $0.33$ \\
$\mathcal{P}$ & HI & $2.5$ & $0.27$ \\
$\mathcal{P}$ & g$\times$HI & $4.7$ & $1.0$ \\
2D$\mathcal{M}$ & g$\times$HI & $5.2$ & $1.2$ \\
3D$\mathcal{M}$ & g & $2.4$ & $0.47$ \\
3D$\mathcal{M}$ & HI & $5.5$ & $2.2$ \\
3D$\mathcal{M}$ & g$\times$HI & $12$ & $3.4$ \\
\bottomrule
\end{tabular}
\captionof{table}{Median Figure of Merit ${\rm FoM}_{m}$ across test sets, for the explored SBI scenarios. Values are normalized by $3.23 \times 10^2$, corresponding to the ${\rm FoM}_{m}$ of $\mathcal{P}^{\rm g \times HI}$ case.}
\label{tab:fom}
\end{minipage}
\hspace{0.011\linewidth}
\begin{minipage}{0.45\textwidth}
\centering
\includegraphics[width=\linewidth]{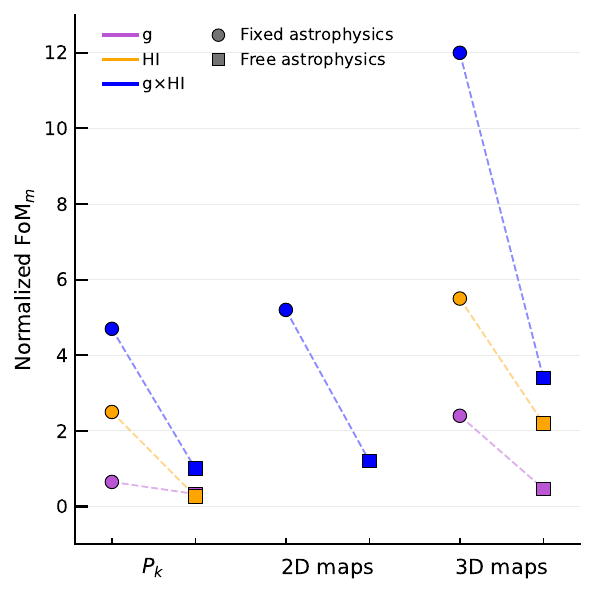}
\caption{Visualization of table \ref{tab:fom} data.}
\label{fig:your_image}
\end{minipage}

\end{figure}

For each combination of tracers and choice of data summary, we quantify the resulting statistical precision on the two cosmological parameters of interest by means of a Figure of Merit (FoM) defined as FoM = $1 / {\rm A}_{68}$, where ${\rm A}_{68}$ is the area within the $68\%$ joint credible region for $\sigma_8$ and $\Omega_{\rm m}$ (marginalized over astrophysical parameters where appropriate). Table \ref{tab:fom} shows the median Figure of Merit (${\rm FoM}_{m}$) computed from the the 200 posteriors in the test set. When moving from $\mathcal{P}^{\rm g \times HI}$ to 3D$\mathcal{M}^{\rm g \times HI}$, ${\rm FoM}_{m}$ increases by a factor $\sim 3$ for both the fixed and free astrophysics configurations, giving a first indication of the gains achievable by a 3D field-level analysis. Comparison with the  2D$\mathcal{M}^{\rm g \times HI}$ cases shows that flatteninig the cubes to 2D maps does not yield comparable increase. 
Furthermore, both in the $\mathcal{P}$ and $\mathcal{M}$ scenarios, the multi-tracers cases are those yielding the highest ${\rm FoM}_{m}$, which increases with respect to the single-tracer cases by factors between $\sim 2$ and up to 7. This demonstrates a non-negligible improvement in the precision achievable by the multi-tracers approach. Accuracy and calibration of the posteriors are addressed below.

\begin{figure}[h!]
    \centering
    \includegraphics[width=\textwidth]{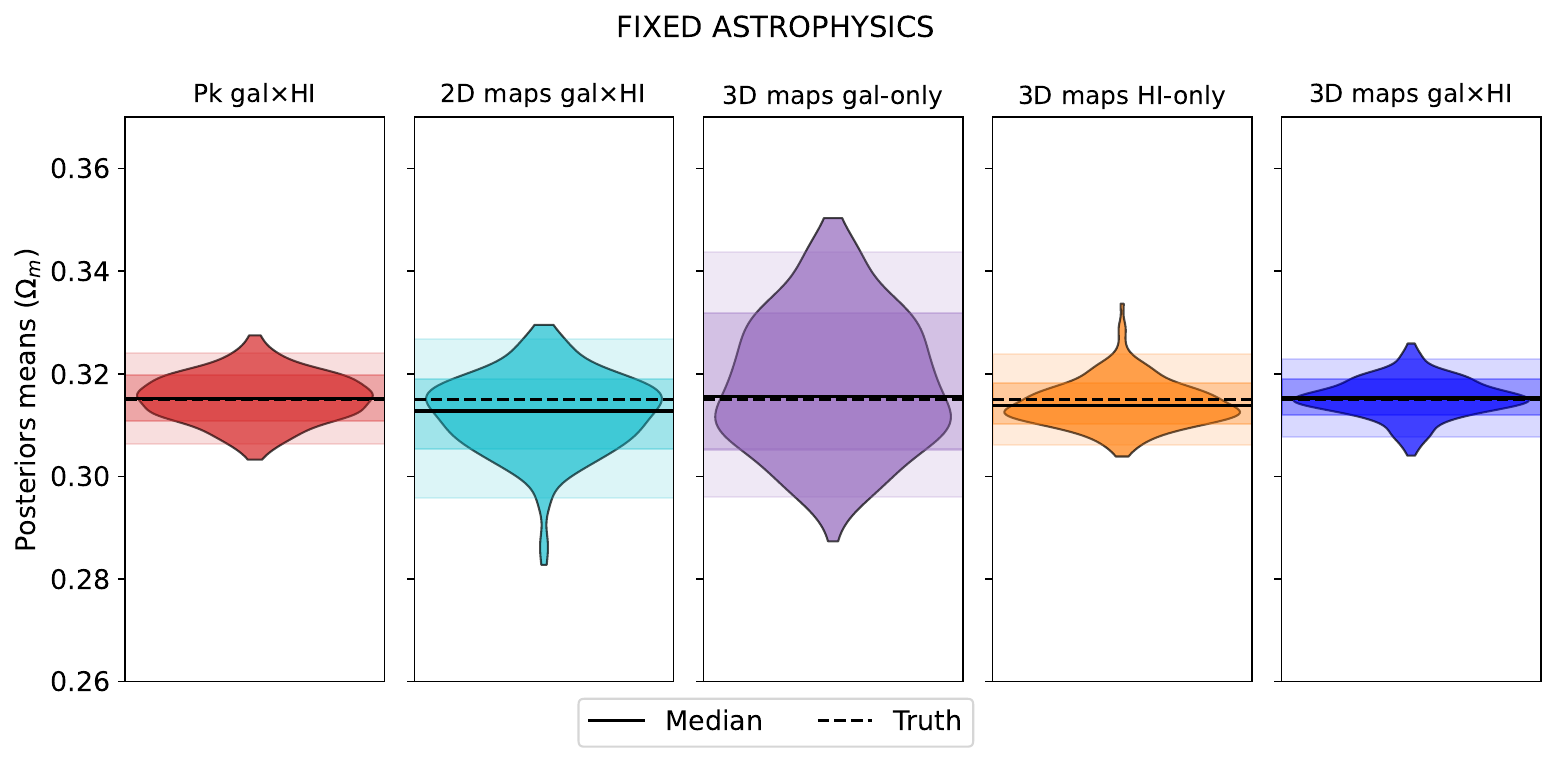}
    
    \vspace{0.5cm}
    
    \includegraphics[width=\textwidth]{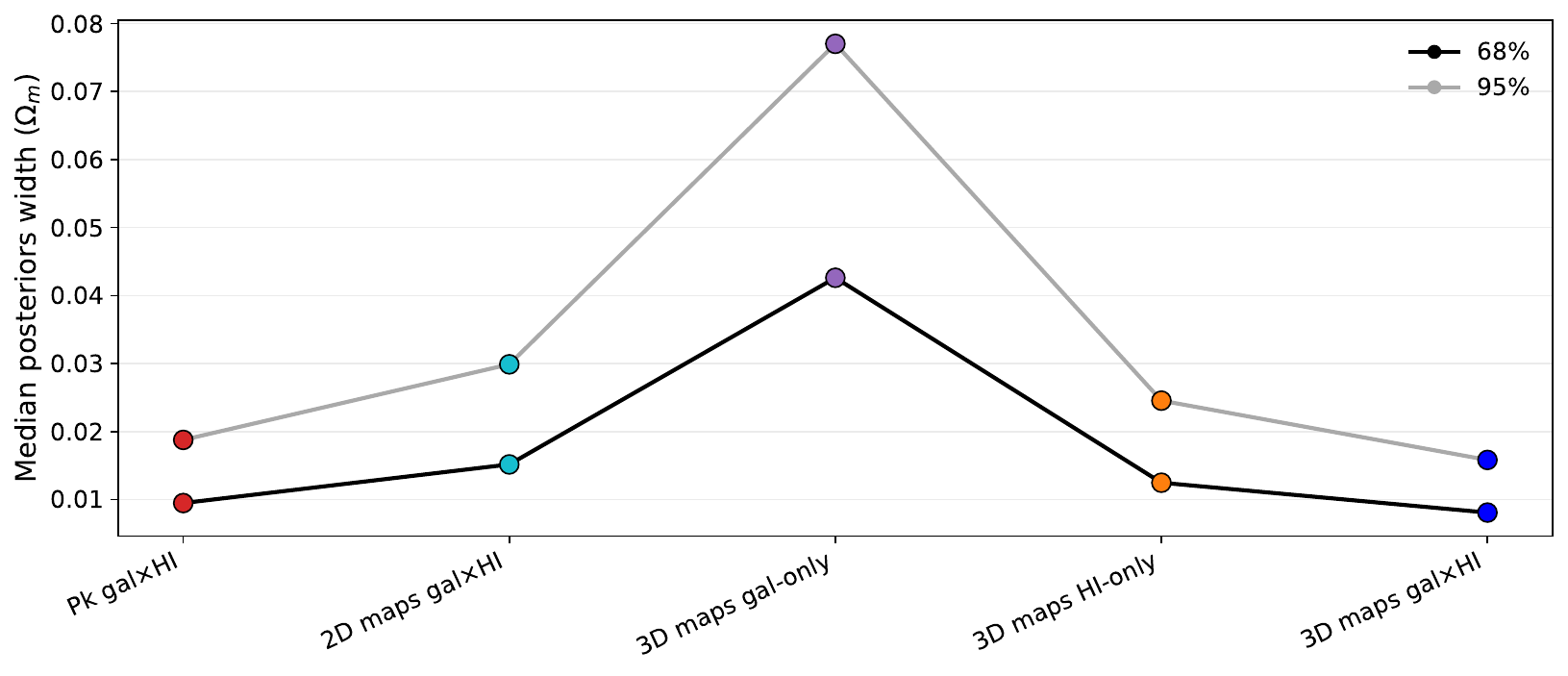}
    \caption{Distribution of marginal posterior summaries for $\Omega_{\rm m}$ (fixed astrophysics case). Top: violin plots showing the distribution of the posterior mean across 200 realization of the $\Lambda$CDM-Planck test set (1,000 samples for each posterior). The distance between the solid line (the median of the all posteriors samples) and the dashed line (the ground truth) can be interpreted as the bias of the inference. Bottom: median value of the marginal posteriors widths of the $\Lambda$CDM-Planck test set for 68-95\% credible intervals; smaller values indicate higher precision levels.}\label{fig:medians_Om_3D_fixedastro}
\end{figure}

\begin{figure}[h!]
    \centering
    \includegraphics[width=\textwidth]{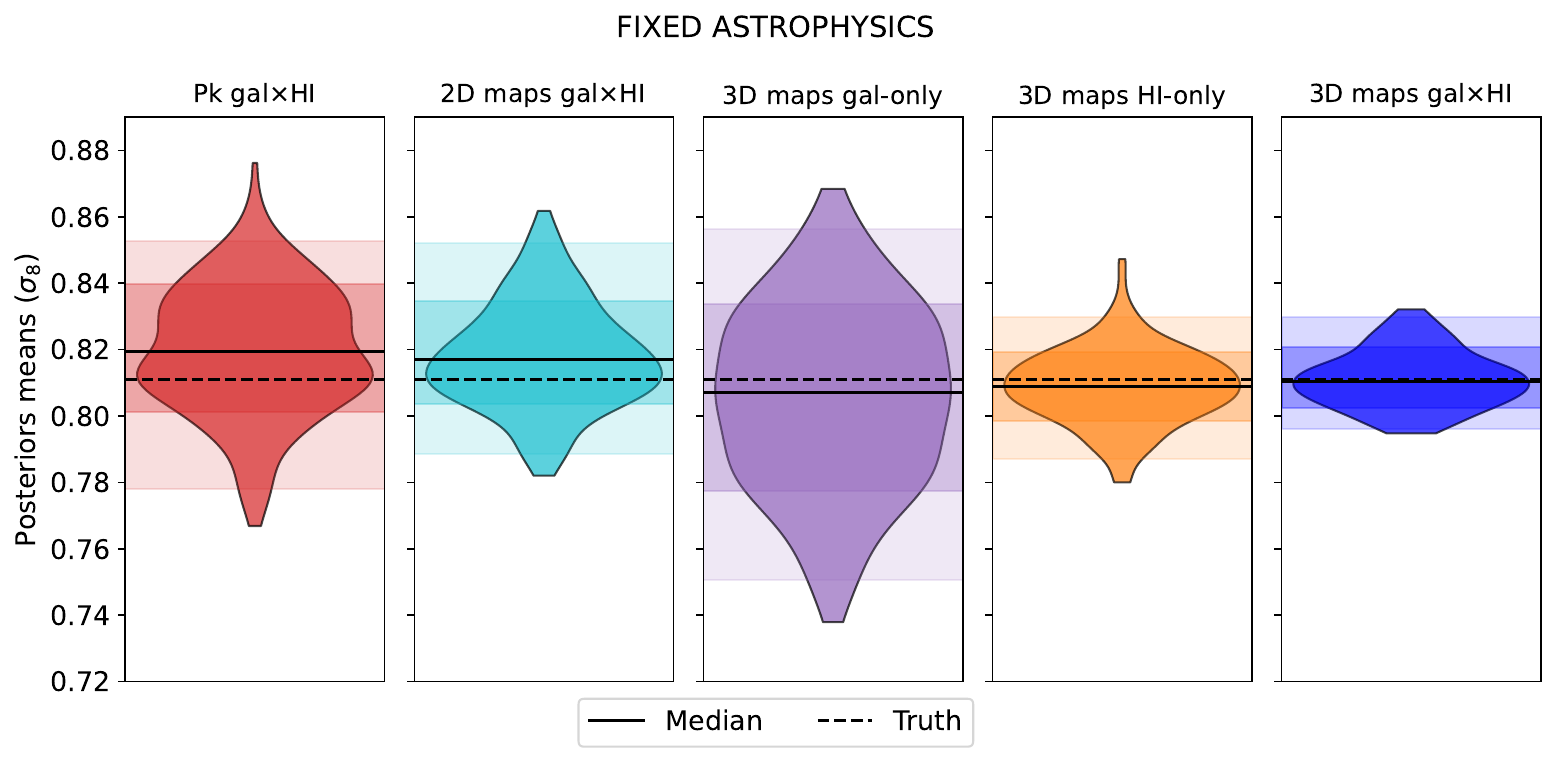}
    
    \vspace{0.5cm}
    
    \includegraphics[width=\textwidth]{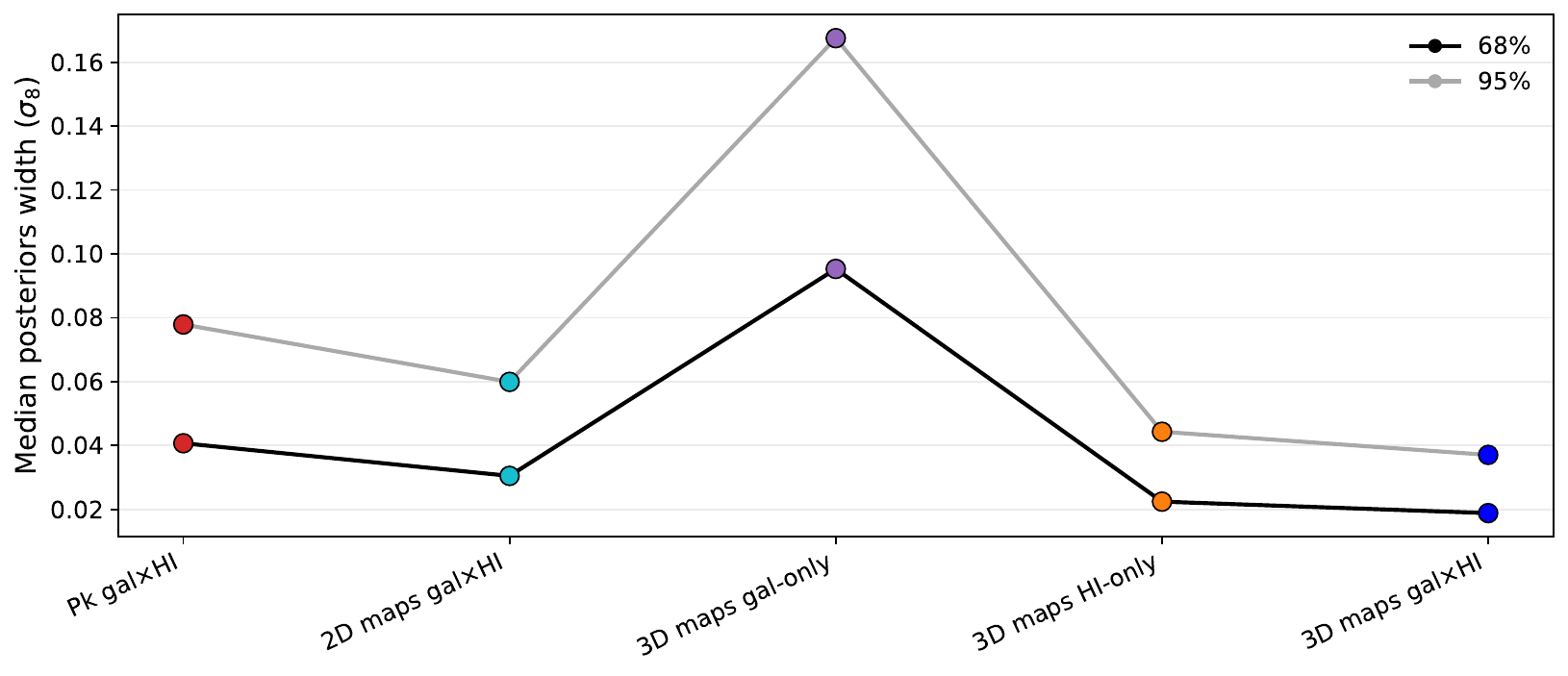}
    \caption{As in figure \ref{fig:medians_Om_3D_fixedastro}, but for $\sigma_8$.}\label{fig:medians_s8_3D_fixedastro}
\end{figure}

\begin{figure}[h!]
    \centering
    \includegraphics[width=\textwidth]{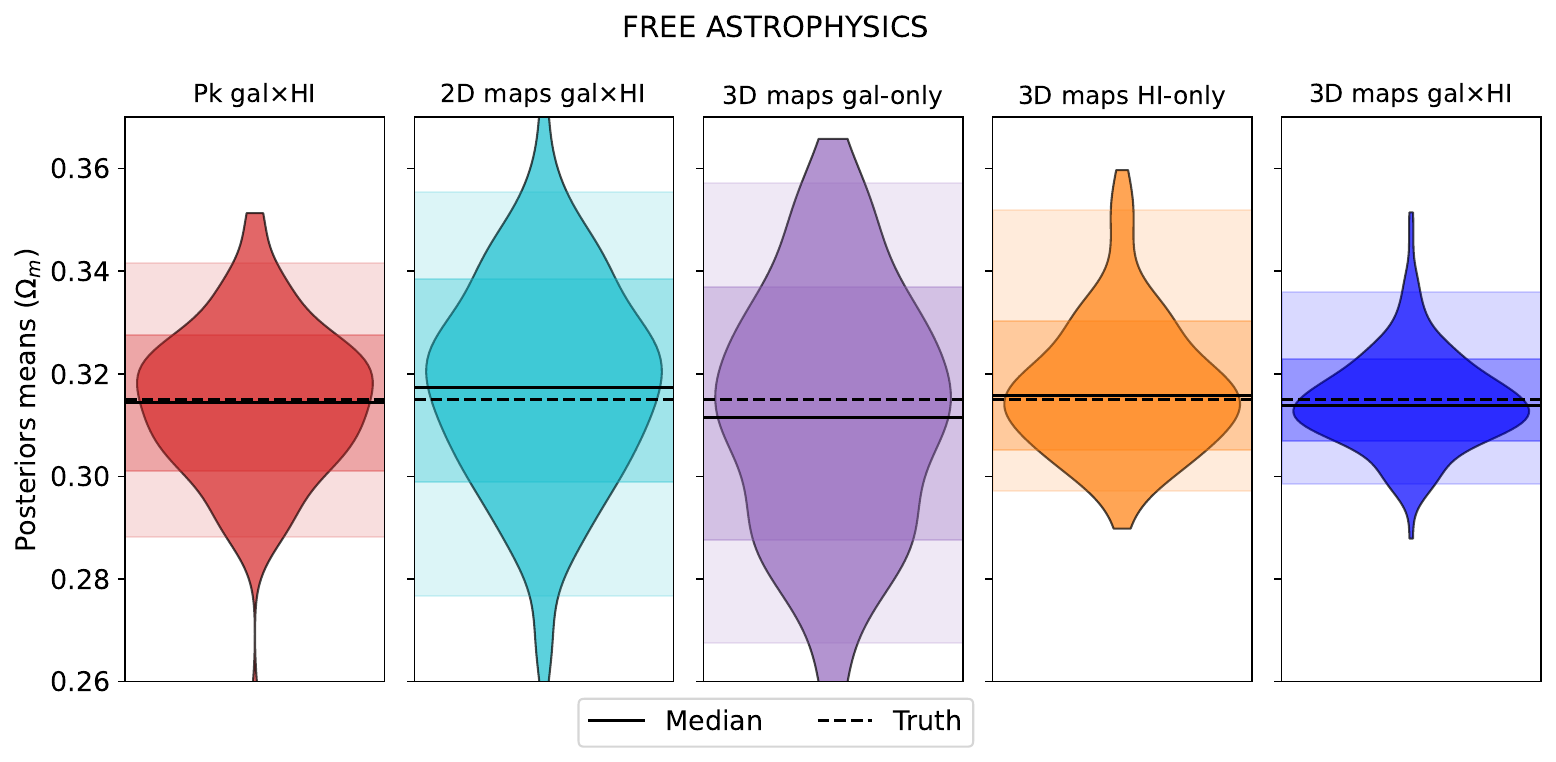}
    
    \vspace{0.5cm}
    
    \includegraphics[width=\textwidth]{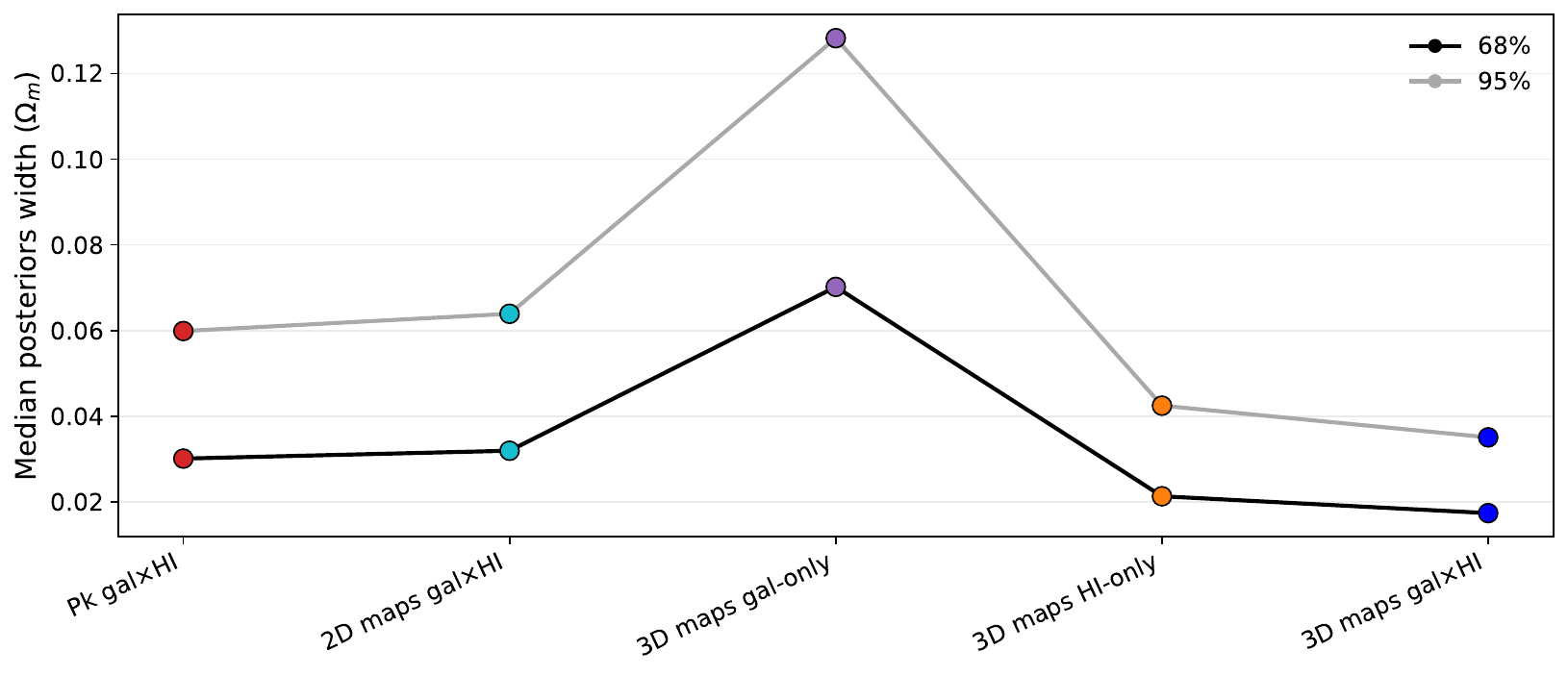}
    \caption{As in figure \ref{fig:medians_Om_3D_fixedastro}, but for the free astrophysics scenario.}\label{fig:medians_Om_3D_freeastro}
\end{figure}

\begin{figure}[h!]
    \centering
    \includegraphics[width=\textwidth]{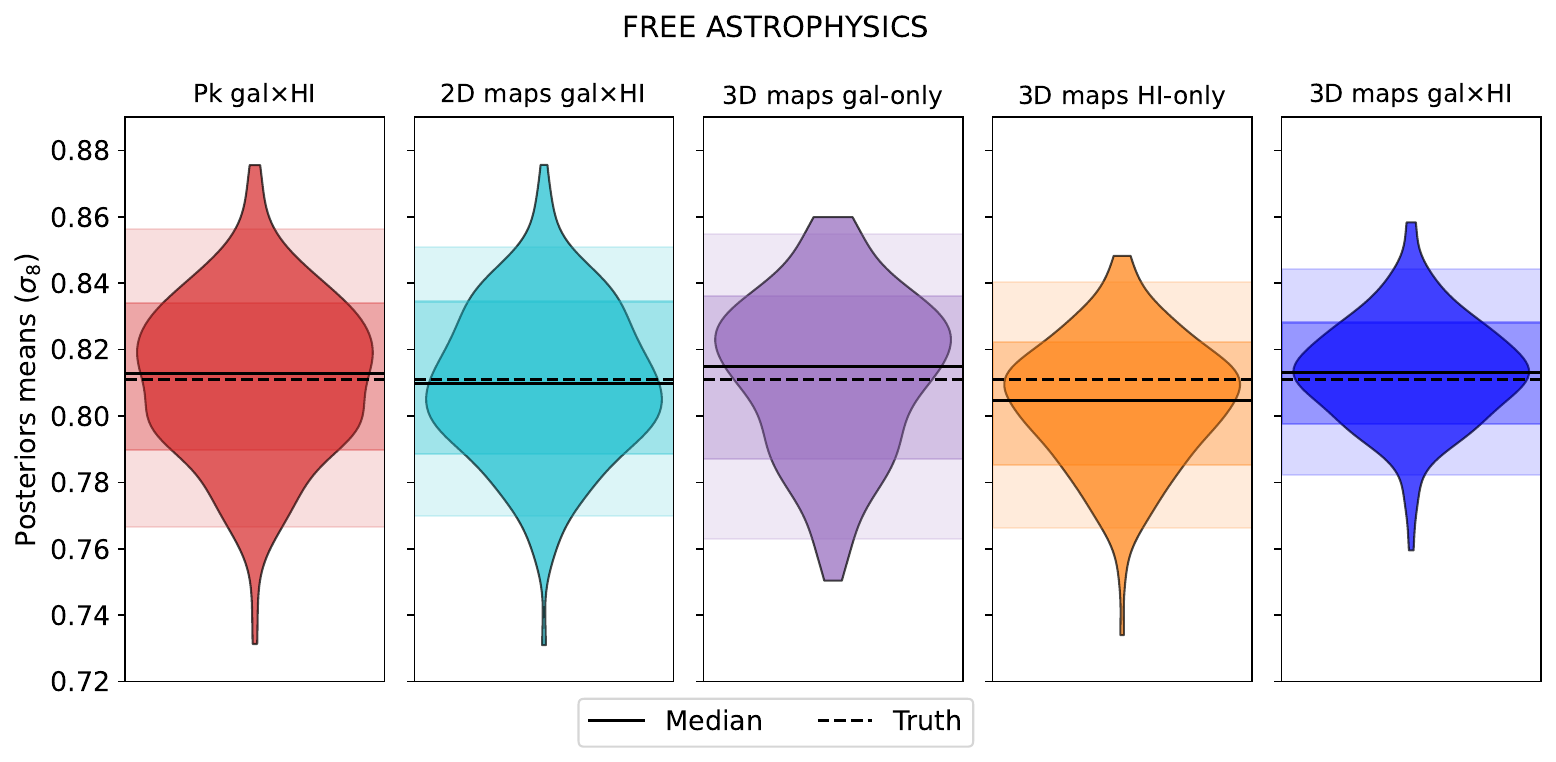}
    
    \vspace{0.5cm}
    
    \includegraphics[width=\textwidth]{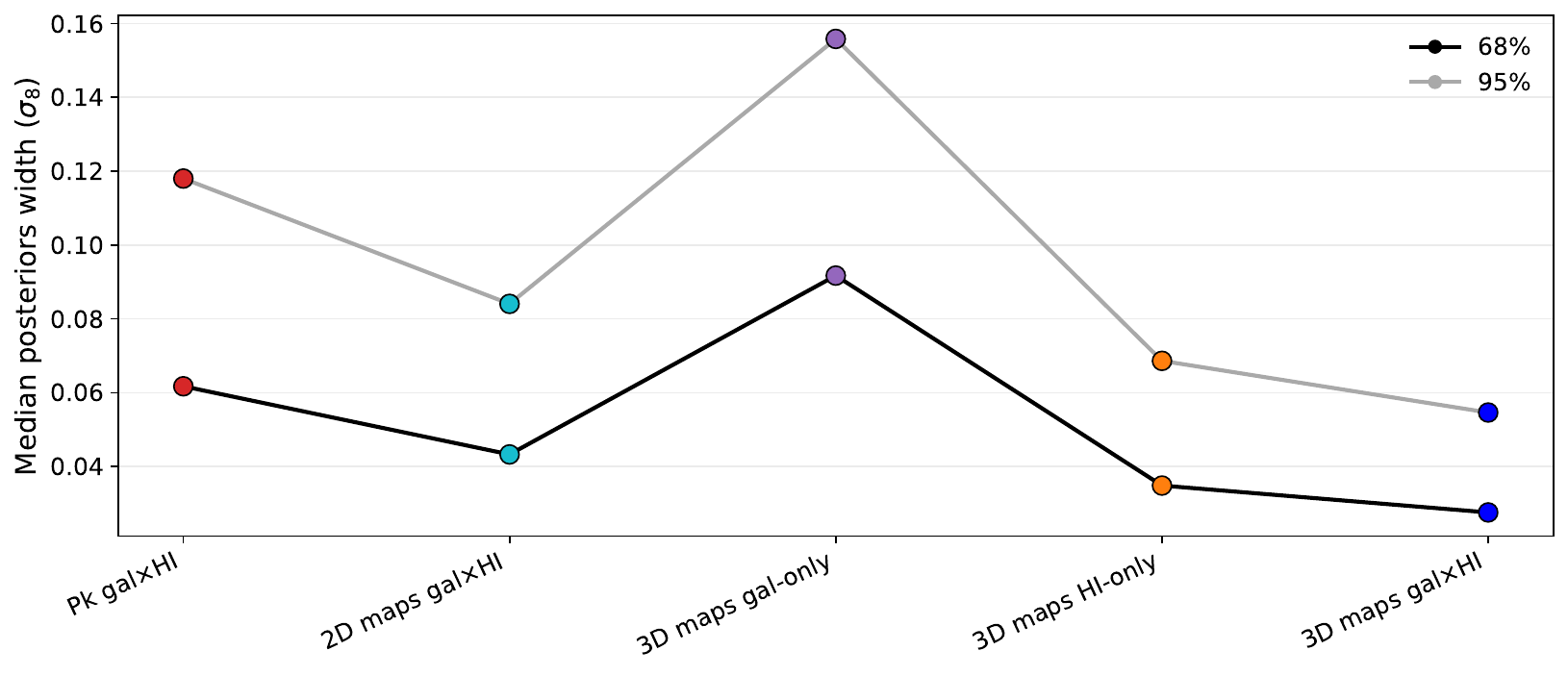}
    \caption{As in figure \ref{fig:medians_Om_3D_freeastro}, but for $\sigma_8$.}\label{fig:medians_s8_3D_freeastro}
\end{figure}

To further investigate the constraining power of the various scenarios, we show in figures \ref{fig:medians_Om_3D_fixedastro}, \ref{fig:medians_s8_3D_fixedastro}, \ref{fig:medians_Om_3D_freeastro} and  \ref{fig:medians_s8_3D_freeastro} the distribution of marginal posterior means over the 200 test realizations (top panel) and their median widths (bottom panel). 
Figures \ref{fig:medians_Om_3D_fixedastro} and \ref{fig:medians_s8_3D_fixedastro} display the results for fixed astrophysics for $\Omega_{\rm m}$ and $\sigma_8$, respectively: the 3D$\mathcal{M}^{\rm g \times HI}_{\rm fixedA}$ case delivers unbiased and narrower posteriors with respect to the $\mathcal{P}^{\rm g \times HI}_{\rm fixedA}$ case. This is especially prominent for $\sigma_8$, whose constraints improve by a factor $\sim 2$ (while $\Omega_{\rm m}$ is improved only mildly).  It is worth noticing that the 2D$\mathcal{M}^{\rm g \times HI}_{\rm fixedA}$ case, instead, does not yield a significant improvement, especially for $\Omega_{\rm m}$, where in fact we observe a degradation of constraints with respect to the $\mathcal{P}^{\rm g \times HI}_{\rm fixedA}$ case. This indicates that, at least under the current setup for the latent representation, the three-dimensionality of the field-level maps is required to retain additional cosmological information.

Focusing the attention on the improvements from combining both tracers, figures \ref{fig:medians_Om_3D_fixedastro} and \ref{fig:medians_s8_3D_fixedastro} also show that posteriors for the 3D$\mathcal{M}^{\rm g \times HI}_{\rm fixedA}$ case are narrower by a factor $\sim 4$ for $\Omega_{\rm m}$ and $\sim 4-5$ for $\sigma_8$ with respect to the 3D$\mathcal{M}^{\rm g}_{\rm fixedA}$ case and narrower by about $\sim 40\%$ for $\Omega_{\rm m}$ with respect to the 3D$\mathcal{M}^{\rm HI}_{\rm fixedA}$ case, indicating a clear advantage from multi-tracer combination. This is most likely due to the fact that galaxy maps are significantly sparser, thereby affected by shot-noise, and built from halos spanning a wide range of masses, each with different bias factors, which effectively mixes objects tracing the underlying DM distribution in different ways. In contrast, HI intensity maps trace the matter field as a smoother continuous observable with a more straightforward dependence on the underlying matter density. Furthermore the relatively small box size used in this study prevents us to fully exploit the capabilities of galaxy clustering at linear scales. The main message of the exercise performed in this work is that the HI field retains more information, at non-linear level, compared to the galaxy field, and our SBI-based framework is able to extract this information.
Indeed, this should not be interpreted as a limitation of galaxy clustering analyses per se, nor as a contradiction with the success of traditional large-scale structure cosmology based on it, since they typically rely on physically motivated bias models (e.g. within the effective field theory framework), large survey volumes, and additional observables such as redshift-space distortions or higher-order statistics, all of which help break degeneracies and recover cosmological information.
A more thorough assessment of the relative merits of each tracer would require consideration of the different noise properties of the observations and systematic effects, whose impact depends on the specific modeling assumptions and observational context, which are beyond the scope of this work. 
Figures \ref{fig:medians_Om_3D_freeastro} and \ref{fig:medians_s8_3D_freeastro} show the same analysis but in the case where astrophysical parameters are marginalized over. In this case, the case 3D$\mathcal{M}^{\rm g \times HI}$ yields less biased posteriors with respect to $\mathcal{P}^{\rm g \times HI}$, and clearly narrower for both cosmological parameters: a $\sim 30\%$ improvement for $\Omega_{\rm m}$ and $\sim 50\%$ for $\sigma_8$, confirming the trends just discussed for the fixed astrophysics scenario.

Using of 200 independent test realizations allows us to assess the stability of the inference across different initial conditions. The dispersion of posterior means (figures \ref{fig:medians_Om_3D_fixedastro}, \ref{fig:medians_s8_3D_fixedastro}, \ref{fig:medians_Om_3D_freeastro}, \ref{fig:medians_s8_3D_freeastro}) indicates that the 3D$\mathcal{M}$ approach not only yields tighter constraints than the $\mathcal{P}$ scenario, but also reduces realization-to-realization scatter, suggesting that this method is less sensitive to cosmic variance, effectively extracting more robust information from each individual map.

\subsection{Bias and calibration}

In order to quantify the bias in the posterior inference of the $\Lambda$CDM-Planck test set, in table \ref{tab:bias} we provide relative percentage values of the distance between the ground truth value and the median value of posterior samples from all realizations in the test set (obtained from 1000 samples for each of the 200 realizations, for fixed $\Lambda$CDM fiducial parameters values), for the scenarios explored in this work. The scenarios presenting the smallest bias are, most of the times, the 3D$\mathcal{M}$ cases. Nonetheless, bias trends are less extreme than the ${\rm FoM}_{m}$ trends reported in table \ref{tab:fom} with bias values never exceeding $\sim 1.5 \%$, well below statistical uncertainties.

\begin{table*}[htbp]
\centering
\renewcommand{\arraystretch}{1.4}
\small

\begin{tabular}{@{}cc|cc|cc@{}}
\toprule
&
&
\multicolumn{2}{c|}{\boldmath$\Omega_m$ \textbf{bias [\%]}} &
\multicolumn{2}{c}{\boldmath$\sigma_8$ \textbf{bias [\%]}} \\
\cmidrule(lr){3-4}
\cmidrule(lr){5-6}
\makecell{\textbf{Data}\\\textbf{summary}} &
\textbf{Tracer} &
\makecell{\textbf{Fixed} \textbf{A.}} &
\makecell{\textbf{Free} \textbf{A.}} &
\makecell{\textbf{Fixed} \textbf{A.}} &
\makecell{\textbf{Free} \textbf{A.}} \\
\midrule
$\mathcal{P}$ & g & 1.5 & 0.98 & 0.21 & 1.0 \\
$\mathcal{P}$ & HI & -0.23 & 1.0 & 0.64 & -0.42 \\
$\mathcal{P}$ & g$\times$HI & 0.042 & -0.19 & 1.0 & 0.23 \\
2D$\mathcal{M}$ & g$\times$HI & -0.72 & 0.75 & 0.73 & -0.15 \\
3D$\mathcal{M}$ & g & 0.19 & -1.1 & -0.49 & 0.48 \\
3D$\mathcal{M}$ & HI & -0.38 & 0.23 & -0.25 & -0.79 \\
3D$\mathcal{M}$ & g$\times$HI & 0.084 & -0.38 & -0.076 & 0.25 \\
\bottomrule
\end{tabular}

\caption{Bias on $\Omega_m$ and $\sigma_8$ across test sets in [\%], for the explored SBI scenarios. We report the percentage difference between ground truth value and median value of the posterior samples across the test set.}
\label{tab:bias}
\end{table*}

Figure \ref{fig:pp-plots_v2_freeastro} displays the same PP-plots for the $\Lambda$CDM-Planck test set under the free astrophysics scenario, where we observe a marked improvement in calibration for the 2D$\mathcal{M}^{\rm g \times HI}$ scenario but a worsening for the 3D$\mathcal{M}^{\rm g \times HI}$ case with respect to the fixed astrophysics setting.

\begin{figure}
    \centering
    \includegraphics[width=0.85\linewidth]{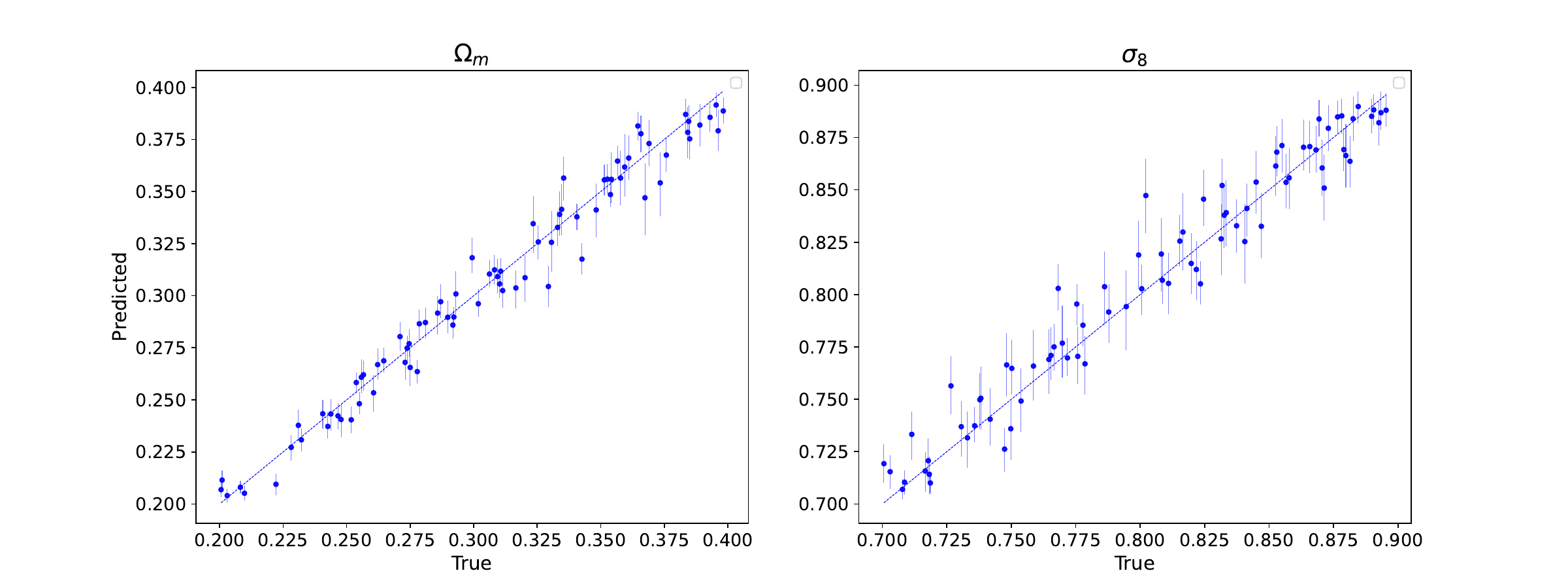}
    \includegraphics[width=0.85\linewidth]{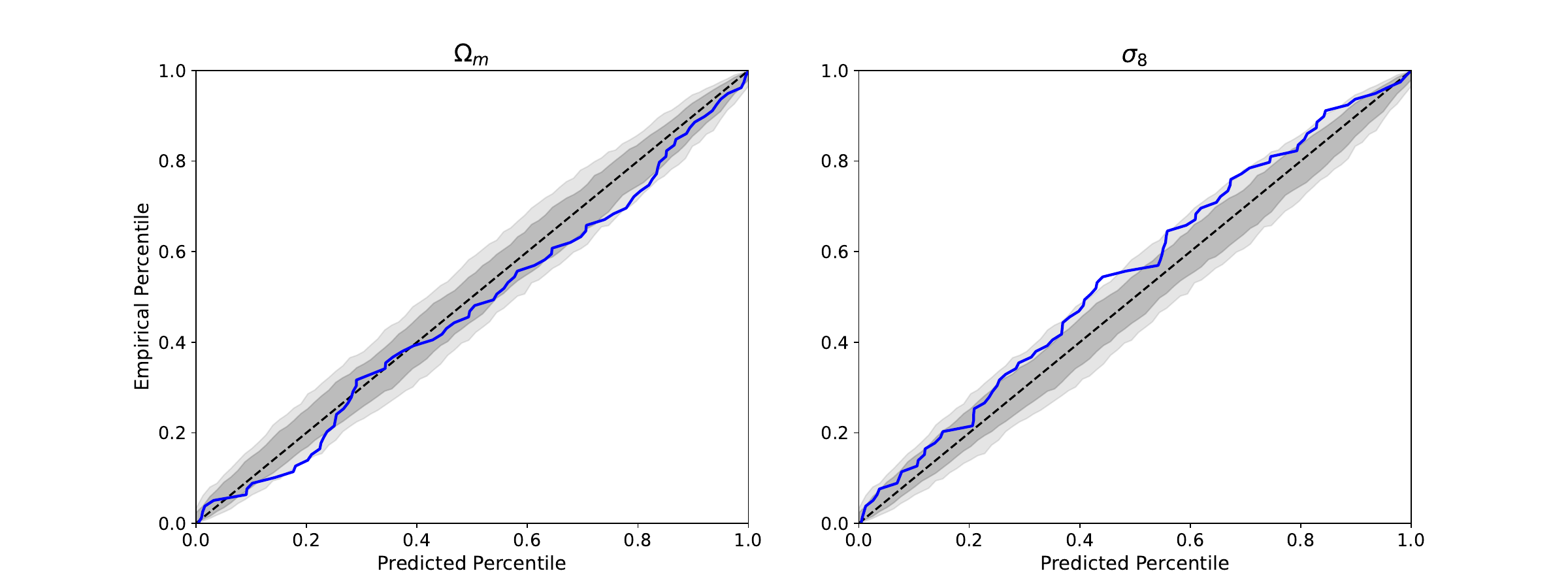}
    \caption{Free astrophysics case 3D$\mathcal{M}^{\rm g \times HI}$, for an emulated test set of 80 maps with true parameters drawn from their priors for both cosmology and astrophysics. \textit{Top:} each blue point shows the median of 1000 (marginal) posterior samples for a single realization, with ground truth value given by the horizontal axis. Error-bars show the 68\% credible interval of the corresponding posterior.
     \textit{Bottom:} PP plots with parameters drawn from their prior ranges (i.e., Bayesian PP plots, averaging over prior distributions).  Deviations within the gray bands are consistent with random sampling noise from a binomial distribution with 68-95\% confidence level. }
    \label{fig:diagonal50}
\end{figure}

Finally, to check the calibration of our SBI pipeline across parameter space, we illustrate, in figure \ref{fig:diagonal50}, the true versus marginally inferred values of cosmological parameters for the free astrophysics case 3D$\mathcal{M}^{\rm g \times HI}$, for value of $\astro$ and $\cosmo$ sampled from their priors.  We also display the PP plots for this same set -- this is a Bayesian PP plot, i.e., showing calibration  when parameters are sampled from their respective priors. We note that the calibration in this case is compatible with the 95\% binomial uncertainty essentially across the whole range, showing good calibration. 

\subsection{Overall remarks}
Overall, these results indicate two main trends. First, the transition from power spectrum-based inference (model $\mathcal{P}$) to field-level representation produces the largest gains in precision when exploiting the full 3-dimensionality of the data (model 3D$\mathcal{M}$). Combining the two tracers at the power spectrum level (model $\mathcal{P}^{\rm g \times HI}$) captures only part of the available information, while the 3D$\mathcal{M}$ configurations recover a substantially larger fraction of the cosmological information, retaining well-calibrated constraining power even when marginalizing over astrophysical uncertainties.  Such gain can be interpreted in terms of the additional information encoded in higher-order, non-Gaussian statistics: while the power spectrum captures only two-point statistics, the latent space representations of the full field retains non-Gaussian features generated by non-linear gravitational evolution and tracers bias. These features are particularly sensitive to $\sigma_8$, which controls the amplitude of fluctuations and therefore the strength of mode coupling. The CNN-based compression is able to extract part of this information, whereas it is irreversibly lost when reducing the data to the power spectrum. The lack of improvement observed in the 2D$\mathcal{M}$ case also suggests that simple dimensionality reduction through projection leads to a substantial loss of information: collapsing along one axis removes line-of-sight structures which appear to carry significant cosmological information. Once again, this highlights that the gain of field-level SBI is specifically due to preserving the full three-dimensional structure, although this comes at the expense of computational cost. Second, moving from single tracer to multi-tracers analysis yields tighter and less biased constraints. This gain can be interpreted as a combination of increased effective signal-to-noise and partial breaking of degeneracies, as the two tracers probe the same underlying matter distribution with different biases and realization noise: the inclusion of cross-correlation information helps disentangle cosmological parameters from tracer-specific effects. 

\section{Discussion and conclusions}\label{sec:conclusions}

We have developed an SBI pipeline combining fast dark matter simulations, neural emulators for galaxy and HI fields, and neural density estimators to perform inference on both the power spectrum and latent representations  of field-level data. This setup allows us to systematically assess the impact of data compression and multi-tracer information on cosmological constraints in an idealized setting.

We find that combining tracers through cross-correlations leads to a clear improvement in constraining power. Joint galaxy–HI analyses consistently outperform single-tracer cases, with a significant reduction of parameter degeneracies. This confirms the advantage of multi-tracer approaches in extracting complementary information from the same underlying matter distribution.
Moving from power spectra to field-level inference yields an additional and substantial gain. While SBI on the power spectrum provides a useful baseline, it does not capture the full information content of the data, especially when marginalizing over astrophysical effects. In contrast, SBI analysis of latent representations of the 3D fields leads to more precise and accurate posteriors, especially for $\sigma_8$. We also find that flattening the 3-dimensional structure of the field-level data into 2-dimensional projected maps does not provide the same gains, thus indicating that the improvement is linked to the three-dimensional structure of the fields. This increased constraining power comes, however, at a significantly higher computational cost.

Astrophysical uncertainties play a critical role. When marginalizing over astrophysical parameters, the constraining power of the power spectra degrades substantially, in some cases leading to uninformative posteriors. By contrast, field-level inference retains a significant fraction of the cosmological information, suggesting that non-Gaussian features help disentangle cosmology from astrophysics.

Our analysis is performed in an idealized proof-of-concept setting, neglecting observational effects such as instrumental noise, survey systematics, and foregrounds. While this allows for a clean assessment of the intrinsic information content, it also means that our current results represent an optimistic, best-case scenario. Incorporating more realistic noise modeling and moving towards survey-like simulations is necessary to assess the applicability of this approach to realistic data. For example, one would have to to apply the beam effect and thermal noise of the instrument for an SKA-like observation to HI maps (e.g. ~\cite{autieri}), and consider the important impact of foregrounds~\cite{carucci25}. Our approach should also be compared with existing likelihood-based pipelines~\cite{berti24}. For application to realistic galaxy surveys, it will be crucial to include selection functions, as well as either spectroscopic or photometric redshift errors to the mock sample. All of these improvements require a dedicated set of simulations on which to re-train our models, and we thus defer this investigation to a future work.

Finally, we emphasize that in this work we limited ourselves to a self-consistent proof-of-concept setup, where both training and inference are performed on emulated maps generated within the same framework. We did not attempt to perform a transfer learning analysis to the different simulation suites (e.g., applying SBI trained on emulated maps to original \texttt{CAMELS} maps or to other suites), since previous studies~\cite{Bayer_2025} have  highlighted the sensitivity of field-level SBI methods to out-of-distribution effects and small-scale discrepancies between simulations, even just for pure N-body codes. This mean that  transfer learning between emulated and simulated maps (and even more, real observations) remains challenging in the absence of dedicated mitigation strategies. Addressing these important issue requires a separate investigation, beyond the scope of the present work, whose goal was instead to establish and validate the proposed field-level multi-tracer SBI framework within the context of a given emulated suited of simulations.

Overall, our results highlight the combined impact of multi-tracer analyses and field-level SBI in unlocking the information content of large-scale structure. Extending these methods to more realistic settings and improving their computational efficiency will be key for their application to upcoming cosmological surveys.

\section*{Acknowledgments}
We thank Daniela Breitman, Gabriella Contardo, Kosio Karchev, Fran\c{c}ois Lanusse, Andre Scaffidi, Christoph Weniger for useful discussions. 
We thank Shy Genel and Francisco Villaescusa-Navarro for sharing with us the second generation 50$\,h^{-1}$Mpc \texttt{CAMELS} suites, from which the \texttt{CV} set was used for figures \ref{fig:emulate_gal50} and \ref{fig:emulate_HI50} of this manuscript.

GS is partially supported by Next Generation EU programme (PNRR). 
SM is supported by the National Recovery and Resilience Plan (PNRR), Dottorati Green/Innovazione under DM 351. SM and MR acknowledge support from the SISSA-Flatiron Exchange Programme. 
RT acknowledges co-funding from Next Generation EU, in the context of the National Recovery and Resilience Plan, Investment PE1 – Project FAIR ``Future Artificial Intelligence Research''. This resource was co-financed by the Next Generation EU [DM 1555 del 11.10.22]. RT and MV are partially supported by the Fondazione ICSC, Spoke 3 ``Astrophysics and Cosmos Observations'', Piano Nazionale di Ripresa e Resilienza Project ID CN00000013 ``Italian Research Center on High-Performance Computing, Big Data and Quantum Computing'' funded by MUR Missione 4 Componente 2 Investimento 1.4: Potenziamento strutture di ricerca e creazione di ``campioni nazionali di R\&S (M4C2-19)'' - Next Generation EU (NGEU). 
GS, MV and RT are also partially supported by INFN INDARK grant. MV is also supported by the INAF Theory Grant ``Cosmological Investigation of the Cosmic Web'' and the SISSA-IDEAS grant. Part of the computations for this work were performed on the Leonardo cluster of CINECA.

\newpage

\newpage

\newpage

\newpage

\newpage

\newpage

\newpage

\newpage

\newpage

\newpage

\begin{figure}[h]
    \centering
    \includegraphics[width=\textwidth]{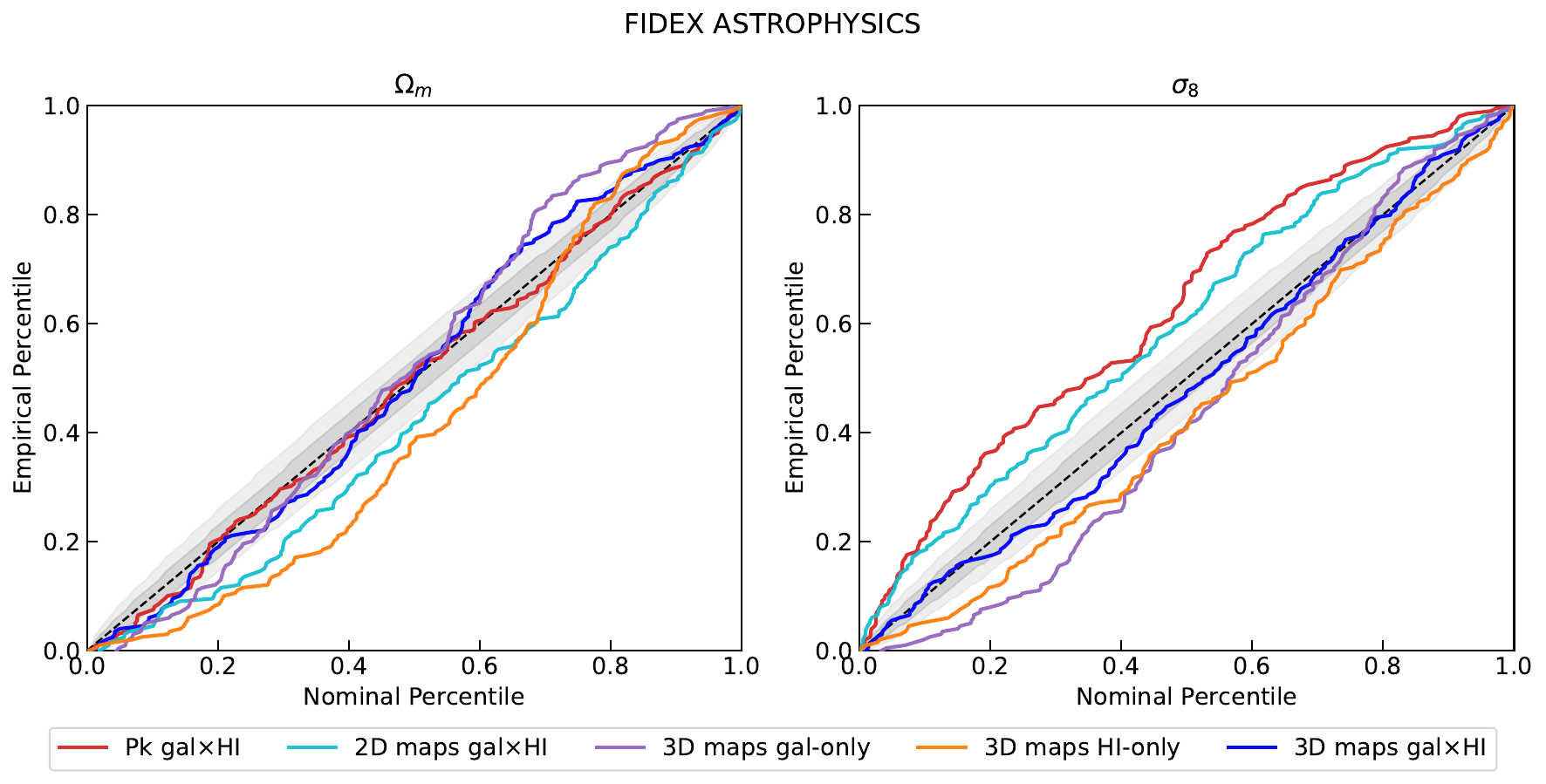}
    \caption{PP plots for the different data configurations for the fixed astrophysics scenario. Deviations within the gray bands are consistent with random sampling noise from a binomial distribution with 68-95\% confidence level. Left panel: $\Omega_{\rm m}$. Right panel: $\sigma_8$.}\label{fig:pp-plots_v2}
\end{figure}
\begin{figure}[h]
    \centering
    \includegraphics[width=\textwidth]{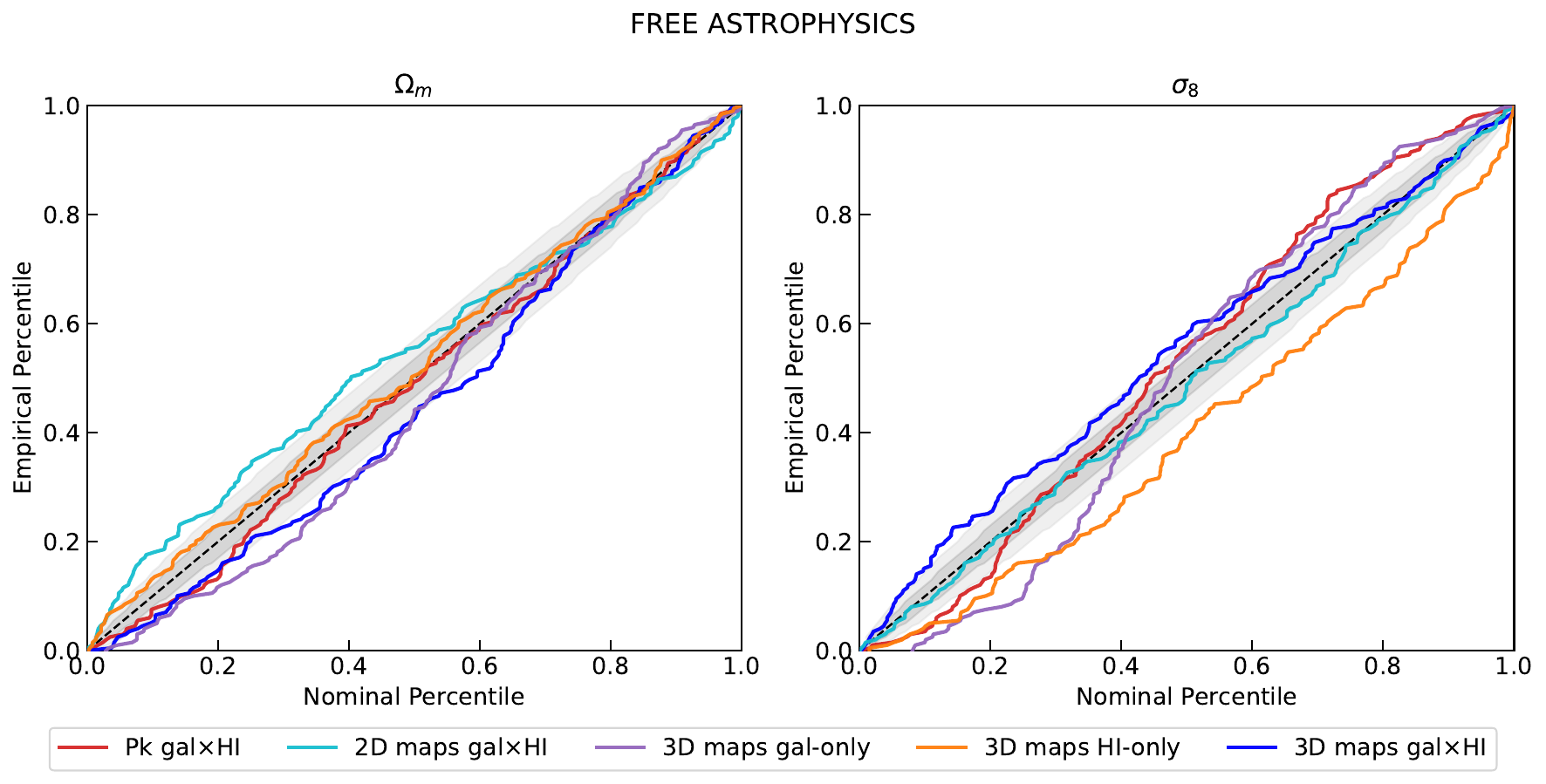}
    \caption{As in figure \ref{fig:pp-plots_v2} but for the free astrophysics scenario.}\label{fig:pp-plots_v2_freeastro}
\end{figure}

\newpage

\newpage

\appendix

\newpage

\section{Emulator}
\label{app:emulator}
Here we describe the neural network-based technique we use as a surrogate model to replace full hydrodynamic simulations, and generate accurate 3 dimensional fields rapidly when trained on reference simulations. 

\subsection{Emulator architecture}

\begin{figure}[tb]
    \centering
    \includegraphics[width=\linewidth]{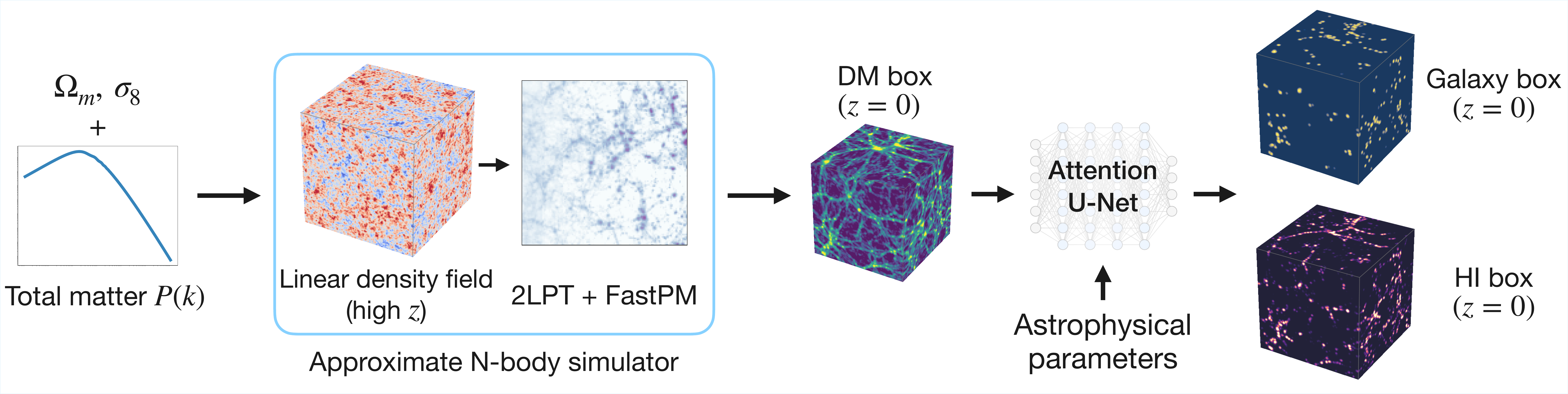}
    \caption{Schematics of the emulation pipeline, from the generation of fast DM maps to the \hgl{} emulator, adapted from~\cite{mishra25}. DM maps, generated by means of \texttt{FastPM}, are passed to the U-net to generate galaxies or HI, conditioning on astrophysics prescriptions. }
    \label{fig:emulator_sketch}
\end{figure}

\begin{table*}
\centering
\caption{Architecture of the \hgl\ emulator. All convolutions (Conv3D) are $3\times3\times3$ with circular padding. Group Normalization (GN) is used throughout, with FiLM conditioning applied within each residual block, as can be seen in the flow diagram in the last row. For stable learning, the second convolutional layer in each ResNet block is initialized to zero.}
\label{tab:architecture}

\begin{tabular}{lccc}
\hline
Stage & Spatial dims & Channels (in $\rightarrow$ out; groups) & Operation \\
\hline

Input & $L^3$ & $1$ & Input field \\

\hline
Encoder & $(L/2)^3$ & $1 \rightarrow 16\ (g{=}1)$ & ResNet block + MaxPool \\
        & $(L/4)^3$ & $16 \rightarrow 32\ (g{=}4)$ & ResNet block + MaxPool \\
        & $(L/8)^3$ & $32 \rightarrow 64\ (g{=}4)$ & ResNet block + MaxPool \\
        & $(L/8)^3$ & $64 \rightarrow 128\ (g{=}4)$ & ResNet block \\

\hline
Bottleneck & $(L/8)^3$ & $128 \rightarrow 128\ (g{=}4)$ & Self-attention \\

\hline
Decoder & $(L/4)^3$ & $128 \rightarrow 64\ (g{=}4)$ & Upsample + ResNet block \\
        & $(L/2)^3$ & $64 \rightarrow 32\ (g{=}4)$ & Upsample + ResNet block \\
        & $L^3$ & $32 \rightarrow 16\ (g{=}4)$ & Upsample + ResNet block \\

\hline
Output & $L^3$ & $16 \rightarrow 1\ (g{=}1)$ & ResNet block \\

\hline
\multicolumn{4}{l}{
\textbf{ResNet block: }
$x \rightarrow \mathrm{GN} \rightarrow \mathrm{FiLM} \rightarrow \mathrm{Conv3D}
\rightarrow \mathrm{GN} \rightarrow \mathrm{FiLM} \rightarrow \mathrm{Conv3D} + x$
}
\\
\hline

\end{tabular}
\end{table*}

The whole emulation pipeline is sketched in figure \ref{fig:emulator_sketch} and the \hgl{} emulator architecture is summarized in table~\ref{tab:architecture}. Following the network architecture of~\cite{mishra25}, the model is based on a 3D U-Net architecture comprising four encoding and four decoding residual convolutional blocks. Each encoding block consists of two 3D convolutional layers, each followed by Group Normalization and a non-linear activation, and is followed by a downsampling operation implemented via max pooling. The decoder mirrors this structure, with each block consisting of two convolutional layers and Group Normalization layers, and spatial resolution is restored via upsampling, using trilinear interpolation. 

Furthermore, to encode the cosmological and astrophysical parameters in this network, we use the FiLM (Feature-wise Linear Modulation) technique~\cite{perez2017filmvisualreasoninggeneral}, which allows to condition the output of the neural network on said parameters. Let $\mathbf{x}\in \mathbb{R}^{C\times D^3}$ denote a feature layer within the residual block with $C$ channels and $D^3$ dimensions, and $\mathbf{c}\in \mathbb{R}^{d_c}$ denote the vector of $d_c$ simulation parameters. Then the FiLM transformation is defined as:
\begin{equation}
    \mathrm{FiLM}(\mathbf{x}, \mathbf{c}) = \boldsymbol{\gamma}(\mathbf{c}) \odot \mathbf{x} + \boldsymbol{\beta}(\mathbf{c}),
\end{equation}
where $\boldsymbol{\beta},\boldsymbol{\gamma}\in \mathbb{R}^C$ are the channel wise scaling and shift parameters of the transformation, and $\odot$ denotes element wise multiplication across channels. In each ResNet block, two of these affine transformations are applied to feature layers, with one pair of $\gamma,\,\beta$ for each channel output at each convolution layer. For example, at the spatial dimension $(L/2)^3$ in the Unet, there are 16 pairs of FiLM transformations, as there are 16 channels in that ResNet block. The FiLM parameters are generated from the conditioning vector via a Multi-Layer Perceptron (MLP) $f_\theta$:
\begin{equation}
    \left[ \boldsymbol{\gamma}(\mathbf{c}), \boldsymbol{\beta}(\mathbf{c}) \right] = f_\theta(\mathbf{c}).
\end{equation}

The conditioning vector $\mathbf{c}$ consists of both cosmological and astrophysical parameters, concatenated into a single vector $\mathbf{c} = [\cosmo,\ \astro] \in \mathbb{R}^{d_c}$ (for our work $\mathbf{c}$ has a dimension of 6). This combined vector is passed through a shared MLP to produce the FiLM modulation parameters.
This design replaces fixed affine transformations in normalization layers with conditioning-dependent transformations, allowing the network to adapt its feature representations dynamically based on the input parameters.

\vspace{\baselineskip}

Given a DM map $\mathcal{F}^{\rm DM}_{i}(\Omega_{m_i}, \sigma_{8_i})$ characterized by a cosmology $i$, the emulator is able to provide the corresponding galaxy/HI map for that same cosmology $i$ and any astrophysics $j$ of choice:
\begin{equation}
\mathcal{F}^{\rm DM}_{i}(\Omega_{m_i}, \sigma_{8_i}) \:\: \xrightarrow[]{\text{U-Net}} \:\: \mathcal{F}^{\rm gal/HI}_{i}(\Omega_{m_i}, \sigma_{8_i}, {\rm A_{SN1_{j}},A_{AGN1_{j}},A_{SN2_{j}},A_{AGN2_{j}}})
\end{equation}
It is important to stress that, although the emulator is able to paint galaxies/HI for any reasonable astrophysical parameters, the cosmology between input and output must stay the same. In other words, the following emulation process is not possible:
\begin{equation}
	\mathcal{F}^{\rm DM}_{i}(\Omega_{m_i}, \sigma_{8_i}) \:\: \xcancel{\xrightarrow[]{\text{\:\:}}} \:\: \mathcal{F}^{\rm gal/HI}_{i}(\Omega_{m_{k\neq i}}, \sigma_{8_{k\neq i}}, {\rm A_{SN1_{j}},A_{AGN1_{j}},A_{SN2_{j}},A_{AGN2_{j}}})
\end{equation}

To preserve spatial information across scales, skip connections are employed by concatenating feature maps from each encoder block to the corresponding decoder block at the same resolution. This design significantly improves reconstruction fidelity by retaining high frequency information lost during downsampling. 

At the bottleneck, again following the previous work, we incorporate a self-attention module that enables the network to capture long range dependencies and non local correlations between voxels in the latent representation, thereby enhancing its ability to model large scale structure.

\subsection{Emulator training}\label{sec:emu_loss}

This section provides details on the emulator training. The input \texttt{FastPM} DM maps (described in section \ref{sec:DM_maps}) of size 25$\,h^{-1}$Mpc are generated for the same ICs seeds of the \texttt{CAMELS} LH suite. The output (target) galaxy/HI maps (described in section \ref{sec:observables}) of same size are instead generated from the corresponding \texttt{CAMELS} hydrodynamical simulations of the same suite.

 The train and validation set maps derive from splitting the 508 different seeds spanning different values of the cosmological (for DM) and cosmological+astrophysical (for galaxies/HI) parameters within the \texttt{CAMELS} priors\footnote{\texttt{CAMELS} priors for LH suite: $\Omega_{\rm m} \in [0.1, 0.5]$, $\sigma_8 \in [0.6, 1.0]$, $A_{\rm SN1} \in [0.25, 4.0]$, $A_{\rm AGN1} \in  [0.25, 4.0]$, $A_{\rm SN2} \in [0.5, 2.0]$, $A_{\rm AGN2} \in [0.5, 2.0]$.}, in such way to have similar distributions of parameters values between train+validation and test sets. To help training, cubes in train+validation sets are repeated for all 24 possible orientations. The train and validation sets combined are then composed by $508 \times 24 = 12,192$ cubes, which are then split in a $75\%$ and $25\%$ train and validation split. The test set is instead composed of 242 cubes without rotations. Train, validation and test sets are composed of only 1 channel, corresponding to the tracer considered. All maps are normalized with standard scaling, fitted on each respective train set.

Following~\cite{mishra25}, the loss function used to train the emulator is a linear combination of multiple Quantile Losses (QL):
\begin{equation}
\mathcal{L} =\sum_{i=1}^{3} \mathrm{QL}(y, \hat{y}; \tau_i)
\end{equation}
where given $y$ and $\hat{y}$ as the target and predicted voxel value, the quantile loss is defined as:
\begin{equation}
\mathrm{QL}(y, \hat{y}; \tau) = 
\frac{1}{N} \sum_{j=1}^{N} 
\max\big( \tau (y_j - \hat{y}_j),\, (\tau - 1)(y_j - \hat{y}_j) \big)
\end{equation}

This allows the model to learn different regions of the conditional distribution of the target field, improving robustness to sparsity. We use three quantile levels given by $\tau_i \in \{0.45,\, 0.60,\, 0.75\}$ for the prediction of HI fields and $\tau_i \in \{0.45,\, 0.60,\, 0.80\}$ for the prediction of galaxy fields. These values are chosen from empirical experience. When training the emulator, we save models checkpoint every time loss is improved. This lets us identify a-posteriori the checkpoint which corresponds to the best prediction of the power spectrum (i.e. lowest residuals between the predicted Pk and the true one) and the best checkpoint in terms of minimum of the loss (i.e. best predicted maps at pixel level). These two checkpoints may not necessary coincide, as the case of the galaxy emulator, whereas we have one single best checkpoint for the HI emulator. Generating a $64^3$ voxels map takes $\sim 10^{-2}$ seconds on an A100 Nvidia GPU.

\begin{figure}
	\centering
	\includegraphics[width=0.70\linewidth]{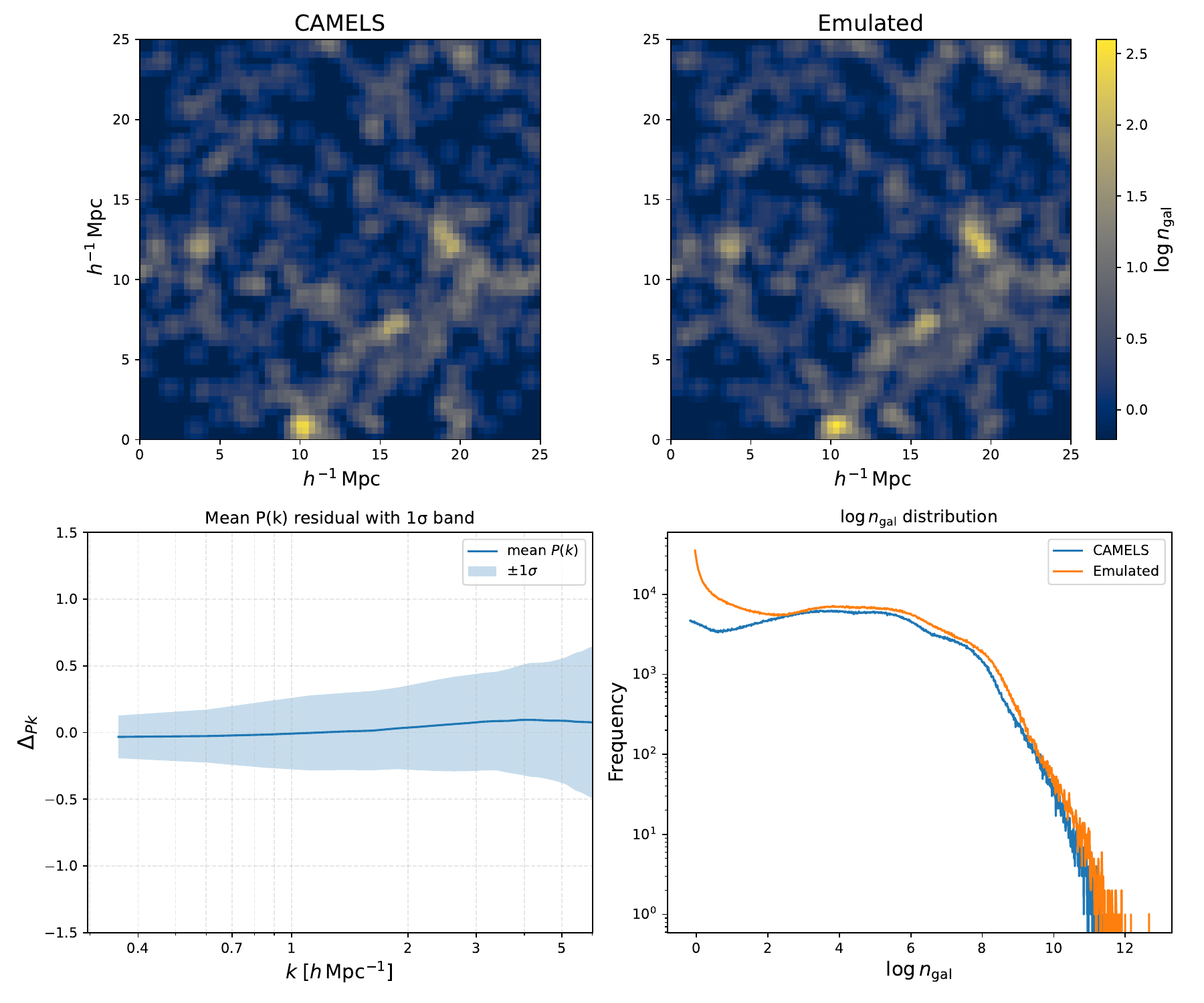}
	\caption{Emulator performance on \texttt{CAMELS} simulations for 25$\,h^{-1}$Mpc box size. The top row compares the \texttt{CAMELS} and emulated galaxy number count fields, for an exemplificative map. The bottom left compares the power spectrum residual $\Delta_\mathrm{Pk}$ between \texttt{CAMELS} from the test set and emulated fields, and the bottom right compares the voxel-wise histogram.}
    \label{fig:emulate_galaxy}
\end{figure}

\begin{figure}
	\centering
	\includegraphics[width=0.70\linewidth]{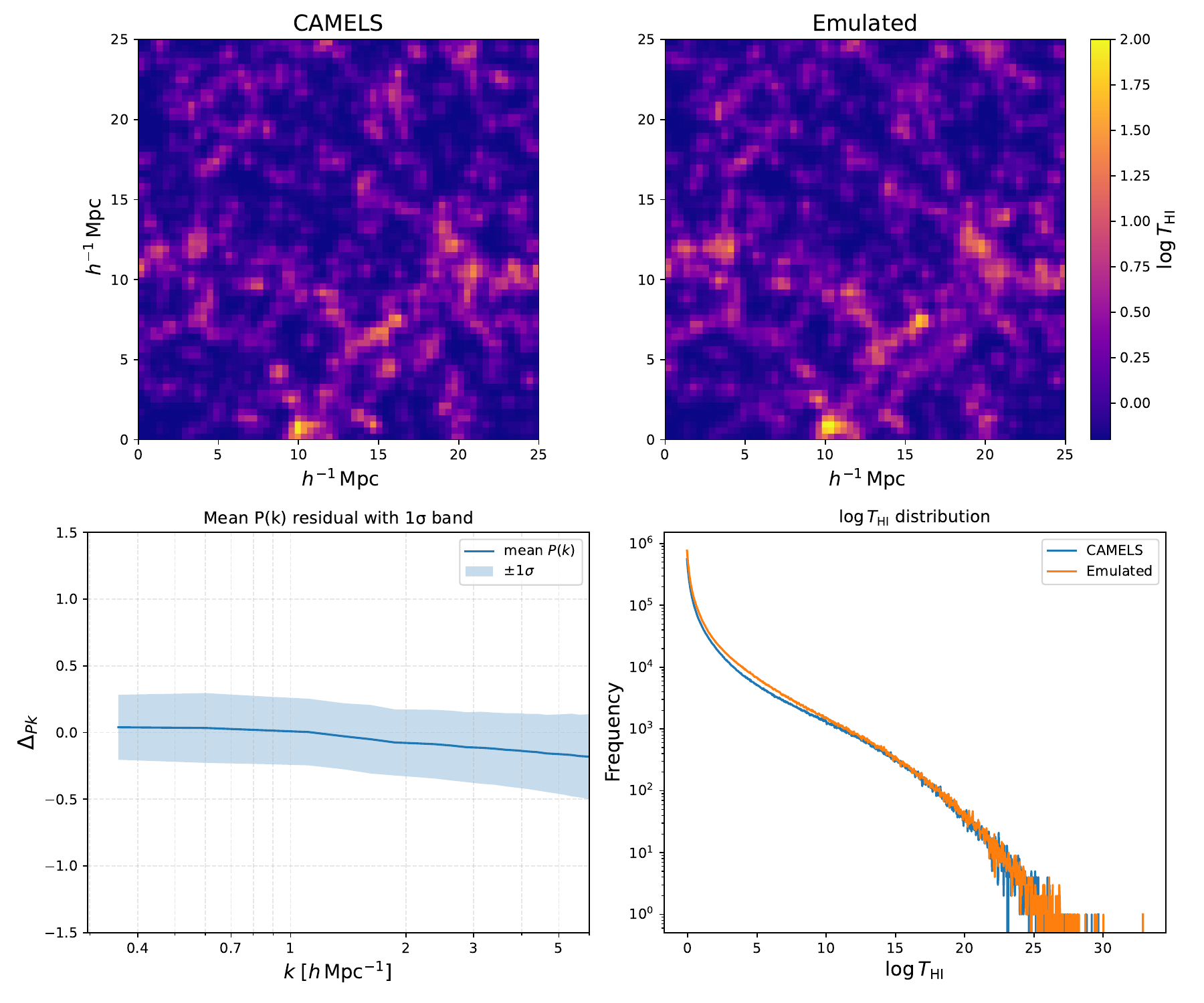}
	\caption{As in figure \ref{fig:emulate_galaxy}, but for the emulated HI temperature fields.}
    \label{fig:emulate_HI}
\end{figure}

Figures \ref{fig:emulate_galaxy} and \ref{fig:emulate_HI} show the emulator performances on 25$\,h^{-1}\rm Mpc$ maps for galaxies and HI respectively. The top row of each figure shows an exemplificative comparison of original \texttt{CAMELS} map from the LH suite and its corresponding emulated counterpart. The bottom row shows, on the left side, the average power spectrum residual $\Delta_{\rm P_k} = ( P_k^{\rm TRUE} - P_k^{\rm EMULATED} ) / P_k^{\rm TRUE}$ [\%] on 242 test maps taken from the LH set, with 1-$\sigma$ shaded band, while on the right side the histograms of pixels values, for original and emulated test maps. Residuals are below 25\% for most of the considered scales with well-centered average values, and good agreement at the histogram level. 

\begin{figure}
	\centering
	\includegraphics[width=0.70\linewidth]{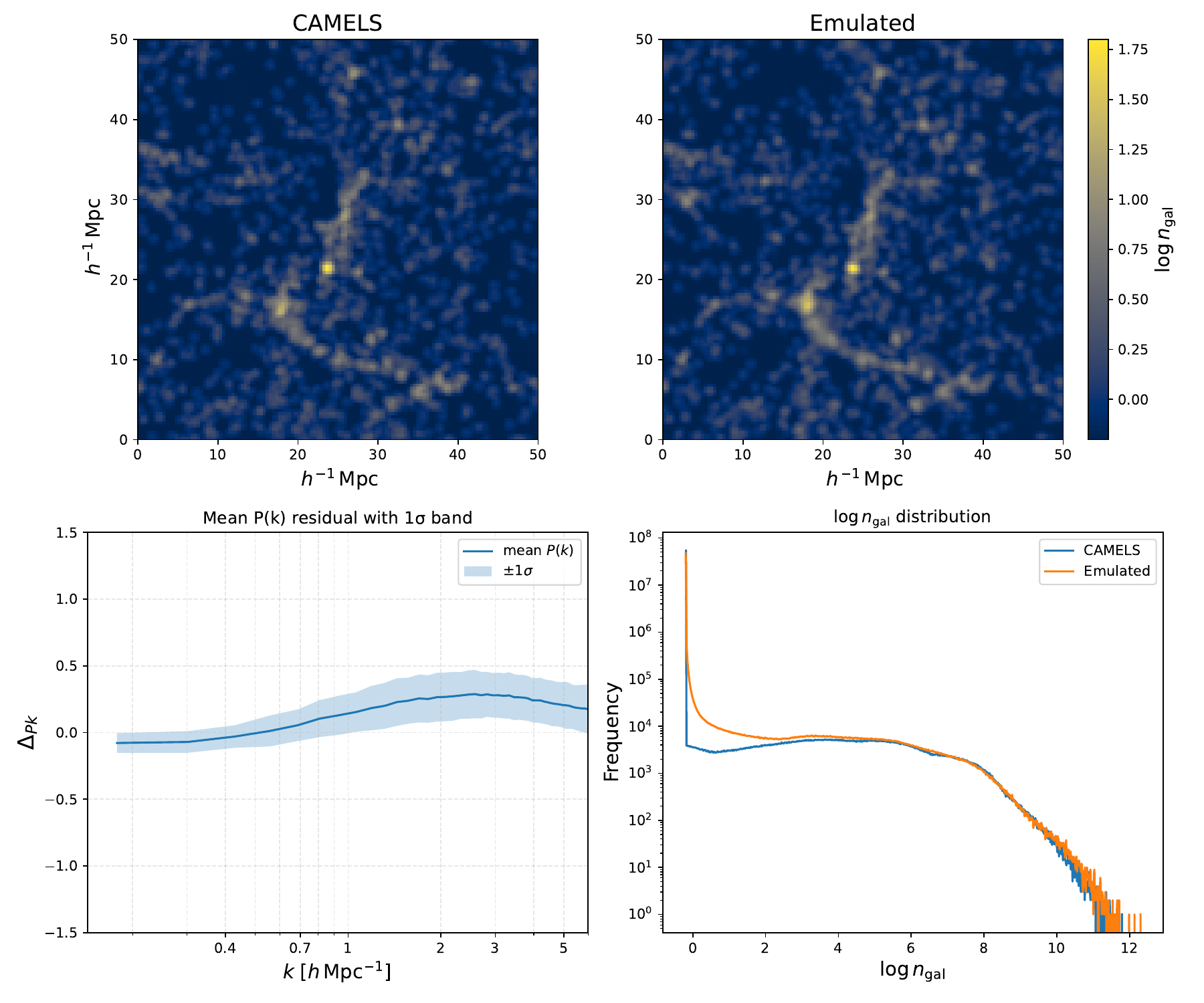}
	\caption{As in figure \ref{fig:emulate_galaxy}, but for (50$\,h^{-1}$Mpc$)^{3}$ maps, generated with the emulator trained on same resolution (25$\,h^{-1}$Mpc$)^{3}$ maps. Test set is the CV \texttt{CAMELS} suite.}
    \label{fig:emulate_gal50}
\end{figure}

\begin{figure}
	\centering
	\includegraphics[width=0.70\linewidth]{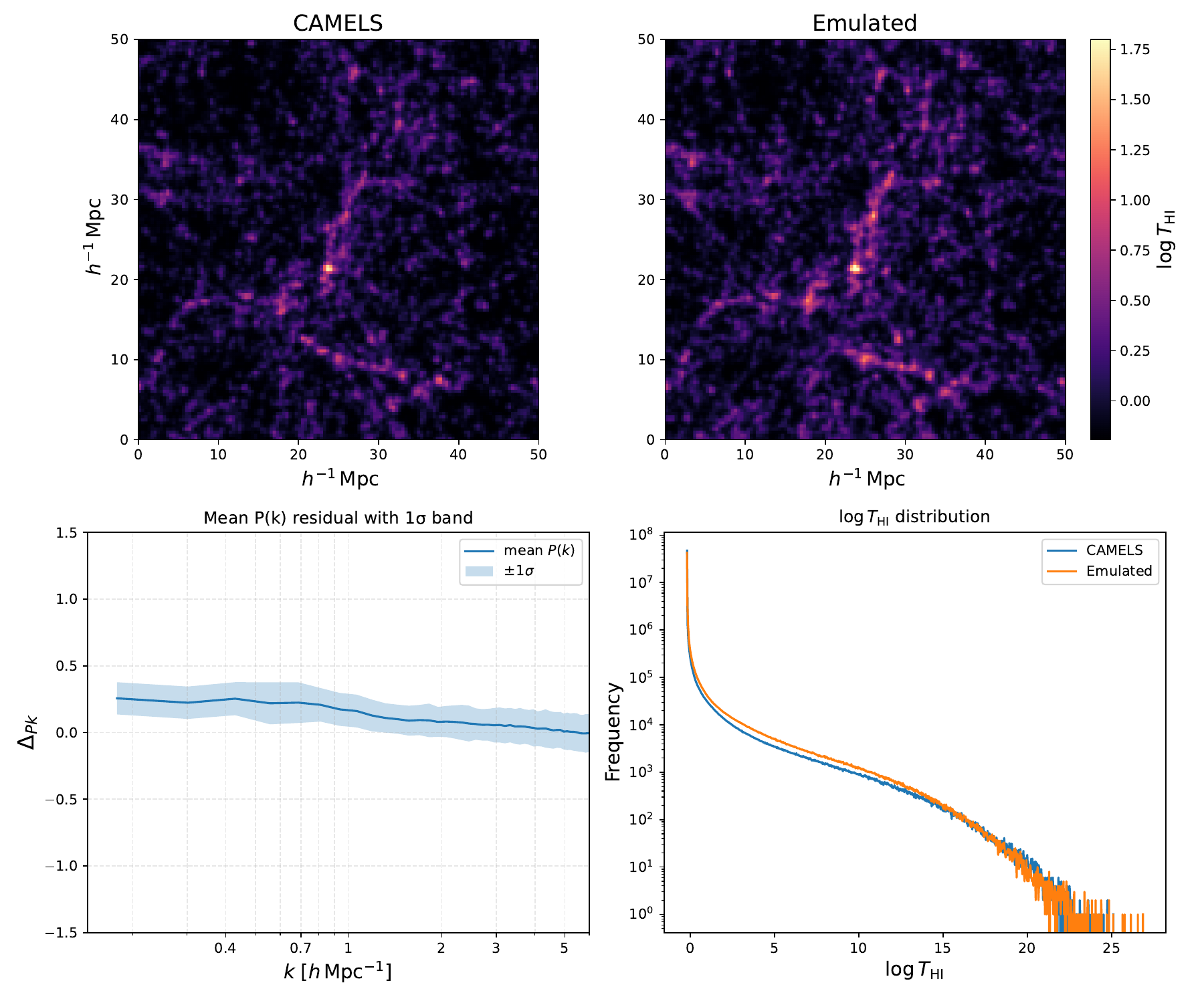}
	\caption{As in figure \ref{fig:emulate_HI}, but for (50$\,h^{-1}$Mpc$)^{3}$ maps, generated with the emulator trained on same resolution (25$\,h^{-1}$Mpc$)^{3}$ maps. Test set is the CV \texttt{CAMELS} suite.}
    \label{fig:emulate_HI50}
\end{figure}

Figures \ref{fig:emulate_gal50} and \ref{fig:emulate_HI50} are the equivalent of figures \ref{fig:emulate_galaxy} and \ref{fig:emulate_HI} but for the 50$\,h^{-1}\rm Mpc$ emulated maps with the same resolution, generated by adopting the same emulator trained on the 25$\,h^{-1}\rm Mpc$ ones, as motivated in section \ref{sec:observables}. Given the absence of the LH set of \texttt{CAMELS} simulations of this larger size, the test set for figures \ref{fig:emulate_gal50} and \ref{fig:emulate_HI50} is given by the CV suite (i.e. 27 seeds at same fiducial cosmology). Although the pixel-values histogram is still well calibrated, the power spectrum appears less accurately emulated than in the 25$\,h^{-1}\rm Mpc$ case. This is likely due to the fact that larger modes (e.g., larger halos) present in the bigger boxes are never seen during training on the smaller box sizes. Nonetheless, given the impossibility to train an emulator on bigger boxes, as explained in section \ref{sec:observables}, we adopt this configuration for the SBI analysis. This is not problematic insofar as we are using test maps generated from the same emulator, as we do here. However, for the future this issue will need to be addressed in order to apply our framework to realistic maps.

\newpage

\bibliography{biblio}
\bibliographystyle{utcaps}

\end{document}